\documentclass[12pt,pdftex]{article}
\pdfoutput=1

\usepackage[a4paper,text={16.8cm,22.4cm}]{geometry}
\usepackage{amsmath,amssymb,bm,graphicx,color}
\allowdisplaybreaks 
\begin{document}
\begin{titlepage}

\begin{flushright}
MZ-TH/12-13\\
March 30, 2012\\
Revised July 20, 2012
\end{flushright}

\vspace{0.2cm}
\begin{center}
\Large\bf
Higgs Production in a Warped Extra Dimension
\end{center}

\vspace{0.2cm}
\begin{center}
Marcela Carena$^{a,b}$, Sandro Casagrande$^c$, Florian Goertz$^d$, \\[1mm] 
Ulrich Haisch$^e$ and Matthias Neubert$^f$\\

\vspace{0.6cm}
{\sl 
${}^a$\,Theoretical Physics Department, Fermilab, Batavia, IL 60510, U.S.A.\\[2mm]
${}^b$\,Enrico Fermi Institute and KICP, University of Chicago,
Chicago, IL 60637, U.S.A.\\[2mm]
${}^c$\,Excellence Cluster Universe, Technische Universit\"at M\"unchen\\ 
D--85748 Garching, Germany\\[2mm]
${}^d$\,Institute for Theoretical Physics, ETH Zurich, 8093 Zurich, Switzerland\\[2mm]
${}^e$\,Rudolf Peierls Centre for Theoretical Physics, University of Oxford\\
OX1 3PN Oxford, United Kingdom\\[2mm]
${}^f$\,PRISMA Cluster of Excellence \& Institut f\"ur Physik (THEP)\\ 
Johannes Gutenberg-Universit\"at, D--55099 Mainz, Germany}
\end{center}

\vspace{0.2cm}
\begin{abstract}
\vspace{0.2cm}
\noindent 
Measurements of the Higgs-boson production cross section at the LHC are an important tool for studying electroweak symmetry breaking at the quantum level, since the main production mechanism $gg\to h$ is loop-suppressed in the Standard Model (SM). Higgs production in extra-dimensional extensions of the SM is sensitive to the Kaluza-Klein (KK) excitations of the quarks, which can be exchanged as virtual particles in the loop. In the context of the minimal Randall-Sundrum (RS) model with bulk fields and a brane-localized Higgs sector, we derive closed analytical expressions for the gluon-gluon fusion process, finding that the effect of the infinite tower of virtual KK states can be described in terms of a simple function of the fundamental (5D) Yukawa matrices. Given a specific RS model, this will allow one to easily constrain the parameter space, once a Higgs signal has been established. We explain that discrepancies between existing calculations of Higgs production in RS models are related to the non-commutativity of two limits: taking the number of KK states to infinity and removing the regulator on the Higgs-boson profile, which is required in an intermediate step to make the relevant overlap integrals well defined. Even though the one-loop $gg\to h$ amplitude is finite in RS scenarios with a brane-localized Higgs sector, it is important to introduce a consistent ultraviolet regulator in order to obtain the correct result.
\end{abstract}
\vfil

\end{titlepage}

\section{Introduction}

In the past decades there has been an enormous effort, both theoretically and experimentally, in trying to understand the origin of electroweak symmetry breaking. Within the Standard Model~(SM), the Higgs mechanism provides a solution that demands the existence of a fundamental self-interacting scalar field, the Higgs boson. After the analysis of significant amounts of data, the LHC and Tevatron experiments seem to observe, for the first time, an excess of events that could be associated with a Higgs-boson signal. If this excess is confirmed, a new set of questions will need to be addressed. In particular, it will be crucial to understand how the electroweak scale $M_{\rm weak}$ is related, if at all, to other scales in nature such as the Planck scale $M_{\rm Pl}$. Searching for a framework that provides a natural connection between these scales has been one of the main reasons to propose theories beyond the SM. Some of these theories, such as supersymmetry, promote a perturbative extension of the SM up to scales close to $M_{\rm Pl}$. Others feature new strong interactions and provide a solution valid only up to energies a few orders of magnitude above the weak scale, without necessarily specifying the ultraviolet (UV) completion of the theory.  

In all cases, new particles are expected to appear at or slightly above the TeV scale to solve the naturalness problem of the Higgs mass parameter. These new particles are expected to have non-negligible effects on the production and decay rates of the Higgs boson. In the event of a Higgs discovery, any departure from SM expectations will then pave the way to probing models of new physics. From this perspective, precision Higgs-boson physics -- much like rare weak decays and flavor-changing neutral current (FCNC) processes -- is an exquisite tool for probing the structure of electroweak interactions at the quantum level. The couplings of the Higgs boson to photons and gluons vanish at tree level in the SM, but they are non-zero at one-loop order and beyond. These couplings are therefore particularly sensitive to new heavy particles, which can propagate in the loops. 

In this article, we concentrate on analyzing Higgs-boson properties in models with a warped extra dimension, so-called Randall-Sundrum (RS) models \cite{Randall:1999ee}, which represent a very appealing alternative to more traditional extensions of the SM, such as supersymmetry. These models provide not only a natural solution to the hierarchy problem, but also a very compelling theory of flavor. They feature a compact extra dimension with a non-factorizable anti-de Sitter~(AdS) metric and two four-dimensional (4D) branes as the boundaries of the warped extra dimension. The AdS background generates an exponential hierarchy of energy scales, so that the fundamental mass scale near the ``UV brane'' is of order $M_{\rm Pl}$, while that near the ``infrared (IR) brane'' is suppressed by an exponential warp factor $e^{-kr\pi}$, with $k$ being the AdS curvature and $r$ the radius of the extra dimension. While all fundamental parameters are of Planck size, $k\sim r^{-1}\sim M_{\rm Pl}$, due to the curvature of the extra dimension the fundamental scale of the theory near the IR brane can be in the range of a few TeV. Hence, as long as the Higgs sector is localized near the IR brane, quantum corrections to the Higgs potential are cut off at the few-TeV scale and the hierarchy problem is solved. In addition, the large hierarchies observed in the spectrum of fermion masses and in the quark mixing matrix arise in a natural way from the localization of bulk fermions in the extra-dimensional space~\cite{Grossman:1999ra,Gherghetta:2000qt,Huber:2000ie}. An important side effect of this mechanism is that flavor-changing couplings between SM fermions, which arise due to the tree-level exchange of Kaluza-Klein~(KK) gauge bosons, are suppressed by the same small overlap integrals that generate the fermion masses \cite{Gherghetta:2000qt}, a scheme referred to as the RS-GIM mechanism \cite{Agashe:2004ay,Agashe:2004cp}.

In order to fulfill the constraints coming from electroweak precision data and FCNC transitions, the low-lying KK excitations of the SM particles must have masses in the range of a few TeV, and hence may be out of the reach for direct production at the LHC. In recent years, many models have been proposed to alleviate these constraints and address the little hierarchy problem of the basic RS model with bulk fermions. The model building becomes rather involved and ranges from considering the effect of brane-localized kinetic terms for the gauge fields \cite{Davoudiasl:2002ua,Carena:2002dz} to enlarging the bulk gauge symmetries \cite{Agashe:2003zs,Agashe:2006at,Bauer:2011ah}, thereby adding a whole new layer of additional matter fields. Another class of models takes the Higgs sector off the IR brane. This is particularly well motivated if the Higgs boson is identified with the $5^{\rm th}$ component of a gauge field (``gauge-Higgs unification'') \cite{Contino:2003ve,Agashe:2004rs}.

In this paper, we will concentrate on studying the effects of KK excitations on the production of a SM-like Higgs boson in the context of the simplest RS model containing bulk fermions and a minimal scalar sector localized on the IR brane.\footnote{Extensions to models with a custodial symmetry in the bulk, or generalized warped extra-dimensional models will be considered elsewhere.} 
In particular, we will scrutinize the gluon-gluon fusion process, for which the presence of KK modes in loop corrections can have important effects. The type of considerations discussed in this work will also be relevant for the Higgs-boson coupling to two photons, which is the other loop-induced coupling of the Higgs field to gauge bosons. At present, the literature contains two sets of results for the computation of Higgs production and decays in RS-type models that are at variance with each other. While in \cite{Casagrande:2010si} the authors find a significant suppression of the gluon-gluon fusion cross section relative to the SM, the authors of \cite{Azatov:2010pf} find an enhancement. Indeed, when applied to the same model, the two groups obtain effects on the $hgg$ amplitude of approximately equal strength but opposite sign. In this article we revisit in detail both calculations and present alternative ways to derive them. For the first time, we obtain closed analytical expressions for both results in terms of the fundamental parameters of the model, valid to all orders in the ratio of the Higgs vacuum expectation value $v$ and the typical mass scale $M_{\rm KK}$ of KK modes. We find that the calculations of \cite{Casagrande:2010si,Azatov:2010pf} are both correct from a mathematical point of view. The origin of their difference is related to a subtlety connected to the fact that, in order to compute its overlap with the fermion wave functions, it is necessary to regularize the Higgs profile in an intermediate step by taking it (slightly) off the IR boundary. The difference between the results obtained by the two groups arises from the different orders in which the Higgs regulator is removed and the sum over the KK tower is performed. We then show that the enhancement in the gluon-gluon fusion cross section obtained in~\cite{Azatov:2010pf} arises from very heavy KK modes. When the model is defined with a (warped) Planck-scale cutoff so as to solve the hierarchy problem, the cross section is suppressed, in accordance with the result obtained in \cite{Casagrande:2010si}.

The paper is organized as follows. In Section~\ref{sec:setup} we define our notation and present the setup of the problem. In Section~\ref{sec:Leff} we present the low-energy Lagrangian for the effective $hgg$ couplings induced by the virtual effects of KK fermions, which is valid for energies below the KK mass scale. We also describe the setup of the two calculations, one summing first over the infinite KK tower and then taking the Higgs regulator to zero, and the other considering the  limits in the reversed order. In Section~\ref{sec:Azatov} we perform the first calculation and reproduce the results of \cite{Azatov:2010pf} in a different way by using the five-dimensional (5D) fermion propagator of the theory. In this way, we generalize the result to all orders in $v/M_{\rm KK}$. In Section~\ref{sec:finitesum} we perform the second calculation, considering first a simplified case, in which it is straightforward to obtain a closed answer for the sum over KK modes. Based on symmetry considerations, we then make a conjecture for how to extend this result to the general case, which reproduces the numerical results obtained in \cite{Casagrande:2010si}. In Section~\ref{sec:interp} we explore in more detail the origin of the difference between the two approaches, which in RS models with an IR-localized Higgs sector arises from unphysical contributions of KK modes with ``trans-Planckian'' masses. We then argue in Section~\ref{sec:cutoff} that the use of a consistent UV regularization scheme, which for instance is provided by the physical UV cutoff inherent in RS models, eliminates these contributions, thus favoring the result corresponding to a suppression of the gluon-gluon fusion cross section. Phenomenological implications of our findings for Higgs-boson searches at the LHC are studied in Section~\ref{sec:pheno}, while Section~\ref{sec:concl} contains our conclusions.

\section{Preliminaries}
\label{sec:setup}

We work with the non-factorizable RS geometry \cite{Randall:1999ee}
\begin{equation}
   ds^2 = e^{-2\sigma(\phi)}\,\eta_{\mu\nu}\,dx^\mu dx^\nu - r^2\,d\phi^2 \,; \qquad 
   \sigma(\phi) = kr|\phi| \,,
\end{equation}
where $x^\mu$ denote the coordinates on the 4D hyper-surfaces of constant $\phi$ with metric $\eta_{\mu\nu}=\mbox{diag}(1,-1,-1,-1)$. The $5^{\rm th}$ dimension is an $S^1/Z_2$ orbifold of size $r$, labeled by $\phi\in[-\pi,\pi]$. Two 3-branes are located at the orbifold fixed points at $\phi=0$ (UV brane) and $\phi=\pi$ (IR brane). The curvature parameter $k$ and the radius $r$ of the extra dimension are assumed to be of Planck size. The warped Planck scale $M_{\rm Pl}\,e^{-\sigma(\phi)}$ sets the effective, position-dependent fundamental scale at a given point along the extra dimension. It serves as a natural UV cutoff, since quantum gravity would become relevant above this scale. In order to solve the (big) hierarchy problem, one assumes that the effective cutoff scale on the IR brane, $\Lambda_{\rm TeV}\equiv M_{\rm Pl}\,e^{-\sigma(\pi)}\equiv M_{\rm Pl}\,\epsilon$, is in the range of 20\,TeV or so. This requires that $L\equiv kr\pi=-\ln\epsilon\approx 34$. The ``little hierarchy problem'', the fact that the Higgs-boson mass is two orders of magnitude smaller than the cutoff scale, is not addressed in the minimal RS framework. The warped curvature scale, $M_{\rm KK}\equiv k e^{-\sigma(\pi)}=k\epsilon$, is assumed to lie somewhat lower, in the range of a few TeV. It sets the mass scale for the low-lying KK excitations of the SM fields. For instance, the masses of the first KK photon and gluon states are approximately $2.45\,M_{\rm KK}$ \cite{Davoudiasl:1999tf}. It will be convenient to introduce a coordinate $t=\epsilon\,e^{\sigma(\phi)}$, which equals $\epsilon\approx 10^{-15}$ on the UV brane and 1 on the IR brane \cite{Grossman:1999ra}. 

We consider the minimal RS model in which all SM matter and gauge fields propagate in the bulk, while the Higgs boson lives on the IR brane. The 5D theory contains three generations of massive fermions $Q=(U,D)^T$ and $u$, $d$, which transform as doublets and singlets under the SM gauge group $SU(2)_L$, respectively. The electroweak gauge symmetry $SU(2)_L\times U(1)_Y$ is broken by the Higgs sector on the IR brane. The 5D fields are functions of $x$ and $\phi$. The doublet fields $Q$ have left-handed zero modes, while the singlet fields $u$, $d$ have right-handed ones. These correspond to the chiral fermions of the SM. 

In this paper we are particularly interested in the couplings of the SM Higgs boson to fermions. In unitary gauge, the Higgs-boson couplings to SM quarks and their KK excitations read \begin{equation}\label{Lhqq}
   {\cal L}_{hqq} = - \sum_{q=u,d} \sum_{m,n}\,g_{mn}^q\,h\,\bar q_L^{(m)} q_R^{(n)}
    + \mbox{h.c.} \,,
\end{equation}
where the Yukawa couplings are given in terms of the overlap integrals \cite{Casagrande:2010si}
\begin{equation}\label{gdef}
\begin{aligned}
   g_{mn}^u 
   &= \frac{\sqrt 2\pi}{L\epsilon} \int_\epsilon^1\!dt\,\delta^\eta(t-1)
    \left[ a_m^{(U)\dagger}\,\bm{C}_m^{(Q)}(t)\,\bm{Y}_u\,\bm{C}_n^{(u)}(t)\,a_n^{(u)}
    + a_m^{(u)\dagger}\,\bm{S}_m^{(u)}(t)\,\bm{Y}_u^\dagger\,\bm{S}_n^{(Q)}(t)\,a_n^{(U)}
    \right] \\
   &= \frac{1}{\sqrt2} \int_\epsilon^1\!dt\,\delta^\eta(t-1)\,\,{\cal U}_L^{\dagger(m)}(t) 
    \bigg( \begin{array}{cc} \bm{0} & \bm{Y}_u \\ 
           \bm{Y}_u^\dagger & \bm{0} \end{array} \bigg)\,{\cal U}_R^{(n)}(t) \,,
\end{aligned}
\end{equation}
and likewise in the down-type quark sector. Here $n$ labels the different mass eigenstates $q^{(n)}(x)$ in the 4D effective theory, such that $n=1,2,3$ refer to the SM up-type quarks $u,c,t$, while $n=4,\dots,9$ label the six fermion modes of the first KK level, and so on. The quantities $\bm{C}_n^{(A)}(t)$ and $\bm{S}_n^{(A)}(t)$ with $A=Q,u,d$ are diagonal $3\times 3$ matrices in flavor space, which contain the $Z_2$-even and odd fermion profiles along the extra dimension, respectively. These can be expressed in terms of combinations of Bessel functions, whose rank depends on the bulk mass parameters $\bm{c}_Q=\bm{M}_Q/k$ and $\bm{c}_{u,d}=-\bm{M}_{u,d}/k$ of the 5D fermion fields \cite{Grossman:1999ra,Gherghetta:2000qt}. The $SU(2)_L$ gauge symmetry in the bulk implies that $SU(2)$-doublet quark fields have common $\bm{c}_Q$ parameters. The 3-component vectors $a_n^{(A)}$, on the other hand, describe the flavor mixings of the 5D interaction eigenstates into the 4D mass eigenstates, which are generated by the Yukawa interactions on the IR brane \cite{Casagrande:2010si}. Because of electroweak symmetry breaking, these vectors are different for $A=U,D,u,d$. For simplicity, from now on we use the generic notation $Q$ for $U,D$ and $q$ for $u,d$. The $3\times 3$ matrices $\bm{Y}_q$ contain the dimensionless Yukawa matrices of the 5D theory, which contrary to the SM are assumed to have an anarchical structure, i.e., they are non-hierarchical matrices with ${\cal O}(1)$ complex elements. The hierarchies of the Yukawa matrices of the SM quarks in the effective 4D theory are explained in terms of a geometrical realization of the Froggatt-Nielsen mechanism in RS models \cite{Huber:2003tu,Csaki:2008zd,Casagrande:2008hr,Blanke:2008zb}. 

In the second line of (\ref{gdef}) we use a compact notation, in which the profile functions of the left-handed (right-handed)  interaction eigenstates that can mix into the left-handed (right-handed) components of the 4D mass eigenstates are collected in 6-component vectors ${\cal Q}_A^{(n)}(t)$. These vectors obey the orthonormality conditions (with $A=L,R$)
\begin{equation}\label{Unorm}
   \int_\epsilon^1\!dt\,{\cal Q}_A^{(m)\dagger}(t)\,{\cal Q}_A^{(n)}(t) 
   = \delta_{mn} \,.
\end{equation}
They are defined in terms of the KK decompositions
\begin{equation}\label{notat}
\begin{aligned}
   \sqrt{r}\,e^{-2\sigma(\phi)} 
    \left( \begin{array}{c} Q_L(x,\phi) \\ q_L(x,\phi) \end{array} \right)
   &= \sum_n \left( \begin{array}{c} \bm{C}_n^{(Q)}(t)\,a_n^{(Q)} \\ 
    \bm{S}_n^{(q)}(t)\,a_n^{(q)} \end{array} \right) q_L^{(n)}(x) 
   \equiv \sqrt{\frac{L\epsilon}{2\pi}}\,\sum_n\,{\cal Q}_L^{(n)}(t)\,q_L^{(n)}(x) \,, \\
   \sqrt{r}\,e^{-2\sigma(\phi)}
    \left( \begin{array}{c} Q_R(x,\phi) \\ q_R(x,\phi) \end{array} \right)
   &= \sum_n \left( \begin{array}{c} \bm{S}_n^{(Q)}(t)\,a_n^{(Q)} \\
    \bm{C}_n^{(q)}(t)\,a_n^{(q)} \end{array} \right) q_R^{(n)}(x) 
   \equiv \sqrt{\frac{L\epsilon}{2\pi}}\,\sum_n\,{\cal Q}_R^{(n)}(t)\,q_R^{(n)}(x) \,,
\end{aligned}
\end{equation}
where $q_{L,R}^{(n)}(x)=\frac12\,(1\mp\gamma_5)\,q^{(n)}(x)$ denote the left- and right-handed components of the $n^{\rm th}$ Dirac fermion.

Note that even the diagonal Yukawa couplings $g_{nn}^q$ derived from (\ref{gdef}) are in general complex numbers, such that pseudo-scalar currents appear in the Lagrangian
\begin{equation}
   {\cal L}_{hqq}^{\rm diag} 
   = - \sum_{q=u,d} \sum_n \left[ \mbox{Re}(g_{nn}^q)\,h\,\bar q^{(n)} q^{(n)} 
    + \mbox{Im}(g_{nn}^q)\,h\,\bar q^{(n)}\,i\gamma_5\,q^{(n)} \right] ,
\end{equation}
which in principle could have interesting implications for phenomenology. Unfortunately, however, for the SM fermions the imaginary parts of the Yukawa couplings are very strongly suppressed \cite{Casagrande:2010si}.

It has been emphasized in \cite{Azatov:2009na} that in order to properly evaluate the Yukawa couplings it is necessary to regularize the Higgs-boson profile. The reason is that with a brane-localized Higgs sector the odd fermion profiles $\bm{S}_n^{(A)}(t)$ are discontinuous at $t=1$, and hence the overlap integral of a product of two such functions with the naive Higgs profile $\delta(t-1)$ is ill defined. In (\ref{gdef}), we have therefore replaced this profile with a regularized delta function $\delta^\eta(t-1)$. The precise shape of this function will be irrelevant; however, it is important that the Higgs profile is non-zero only within a small interval of width $\eta\ll 1$ next to the IR brane, $\delta^\eta(t-1)\ne 0$ only if $1-\eta<t<1$, and that it has unit area. The $\eta$-dependence of the Yukawa couplings is implicit in our notation, but it will play an important role in our analysis.

In the presence of the regularized Higgs profile, the bulk equations of motion (EOMs) for the profile functions read \cite{Casagrande:2010si}
\begin{equation}\label{eoms}
\begin{aligned}
   \frac{d}{dt}\,{\cal Q}_L^{(n)}(t) 
   &= -x_n\,{\cal Q}_R^{(n)}(t) + {\cal M}_q(t)\,{\cal Q}_L^{(n)}(t) \,, \\
   - \frac{d}{dt}\,{\cal Q}_R^{(n)}(t) 
   &= -x_n\,{\cal Q}_L^{(n)}(t) + {\cal M}_q(t)\,{\cal Q}_R^{(n)}(t) \,,
\end{aligned}
\end{equation}
where $x_n=m_{q_n}/M_{\rm KK}$ are the mass eigenvalues, and
\begin{equation}\label{Mqdef}
   {\cal M}_q(t) 
   = \frac{1}{t}\,\bigg( \begin{array}{cc} \bm{c}_Q & \bm{0} \\ 
                         \bm{0} & -\bm{c}_q \end{array} \bigg) 
    + \frac{v}{\sqrt 2 M_{\rm KK}}\,\delta^\eta(t-1)\,
    \bigg( \begin{array}{cc} \bm{0} & \bm{Y}_q \\ 
           \bm{Y}_q^\dagger & \bm{0} \end{array} \bigg) 
\end{equation}
is the generalized mass matrix. Without loss of generality, the hermitian bulk mass matrices $\bm{c}_A$ of the 5D theory can be taken to be diagonal. The boundary conditions are such that the odd profiles vanish on the two branes, which implies
\begin{equation}\label{IRbcs}
   ( \bm{0} \quad \bm{1} )\,{\cal Q}_L^{(n)}(t_i) = 0 \,, \qquad
   ( \bm{1} \quad \bm{0} )\,{\cal Q}_R^{(n)}(t_i) = 0 \,; \qquad 
   \mbox{for} \quad t_i=\{\epsilon,1\} \,.
\end{equation}
Note that these simple Dirichlet boundary conditions only hold because the Higgs profile has been regularized and is no longer singular on the IR brane. When $t$ is lowered away from 1, the profile functions change rapidly over the small interval in which the Higgs profile is non-zero. A careful analysis of the solutions in the vicinity of the IR brane shows that at $t=1-\eta$ the solutions obey the equation (assuming $\eta\ll 1$) \cite{Casagrande:2010si}
\begin{equation}\label{naiveBCs}
   \bigg( \frac{v}{\sqrt 2 M_{\rm KK}}\,\bm{\tilde Y}_q^\dagger ~\quad \bm{1} \bigg)\,
    {\cal Q}_L^{(n)}(1-\eta) = 0 \,, \qquad
   \bigg( \bm{1} ~\quad - \frac{v}{\sqrt 2 M_{\rm KK}}\,\bm{\tilde Y}_q \bigg)\,
    {\cal Q}_R^{(n)}(1-\eta) = 0 \,,
\end{equation}
where 
\begin{equation}\label{Xqdef}
   \bm{\tilde Y}_q = \frac{\tanh\bm{X}_q}{\bm{X}_q}\,\bm{Y}_q \,, 
   \qquad \mbox{and} \quad
   \bm{X}_q = \frac{v}{\sqrt2 M_{\rm KK}}\,\sqrt{\bm{Y}_q\,\bm{Y}_q^\dagger}
\end{equation}
is a positive definite, hermitian matrix given in terms of the 5D anarchic Yukawa matrices of the model. Note that the mixed boundary conditions (\ref{naiveBCs}) hold irrespective of the shape of the regularized Higgs profile, as long as it is confined to a box of width $\eta$. It is thus possible to take the limit $\eta\to 0$ corresponding to a brane-localized Higgs boson, drop the Yukawa couplings in the bulk EOMs (\ref{eoms}) and (\ref{Mqdef}), and impose the mixed boundary conditions on the IR brane, i.e., at $t=1^-$, where $1^-$ denotes a point infinitesimally to the left of the IR brane.\footnote{For $\eta\to 0$ the profile functions are discontinuous on the IR brane, so it is necessary to stay an infinitesimal amount away from the boundary.} Note, however, that to derive the result (\ref{naiveBCs}) it was important to regularize the Higgs profile in an intermediate step \cite{Azatov:2009na}. 

The mixed boundary conditions (\ref{naiveBCs}) evaluated at $t=1^-$ are sufficient to determine the masses and profiles of the fermionic KK modes required, e.g., for the analysis of many tree-level flavor-changing processes \cite{Casagrande:2008hr}. However, when computing overlap integrals of fermion profiles with the wave function of the Higgs boson the correct procedure is to start from a regularized Higgs profile, compute the relevant overlap integrals, and then take $\eta$ to zero \cite{Azatov:2009na}. Using the explicit expressions for the profile functions near the IR brane derived in \cite{Casagrande:2010si}, which were obtained from the solution of the system of equations (\ref{eoms}) to (\ref{IRbcs}) assuming $\eta\ll 1$, it is straightforward to derive from (\ref{gdef}) that
\begin{equation}
\begin{aligned}
   g_{mn}^q 
   &= \frac{\sqrt 2\pi}{L\epsilon} \int_\epsilon^1\!dt\,\delta^\eta(t-1)\,
    a_m^{(Q)\dagger}\,\bm{C}_m^{(Q)}(1-\eta)\,\frac{1}{\cosh^2\!\bm{X}_q} \\
   &\quad\times \bigg[ \cosh^2\!\left( \bar\theta^\eta(t-1)\,\bm{X}_q \right) 
    - \sinh^2\!\left( \bar\theta^\eta(t-1)\,\bm{X}_q \right) \bigg]\,
    \bm{Y}_q\,\bm{C}_n^{(q)}(1-\eta)\,a_n^{(q)} \,,
\end{aligned}
\end{equation}
where the cosh and sinh functions in brackets arise from the $Z_2$-even and odd fermion profiles in (\ref{gdef}), respectively. We have introduced the notation
\begin{equation}\label{thetadef}
   \bar\theta^\eta(t-1) = \int_t^1\!ds\,\delta^\eta(s-1) 
\end{equation}
for the integral of the regularized Higgs profile, such that $\bar\theta^\eta(0)=0$ and $\bar\theta^\eta(t-1)=1$ for $t\le 1-\eta$. We observe that the products of the $t$-dependent fermion profiles add up to a constant, so we are left with the normalization integral over the Higgs-boson wave function, which equals 1. We thus obtain
\begin{equation}\label{g1mi}
   \lim_{\eta\to 0}\,g_{mn}^q 
   = \frac{\sqrt 2\pi}{L\epsilon}\,
    a_m^{(Q)\dagger}\,\bm{C}_m^{(Q)}(1^-)\,\frac{1}{\cosh^2\!\bm{X}_q}\,\bm{Y}_q\,
    \bm{C}_n^{(q)}(1^-)\,a_n^{(q)} \,.
\end{equation}

Interestingly (and somewhat unexpectedly), the same result can be obtained in a more naive way. By evaluating relation (\ref{gdef}) with a brane-localized Higgs boson, we obtain
\begin{equation}
\begin{aligned}
   \lim_{\eta\to 0}\,g_{mn}^q 
   &= \frac{\sqrt 2\pi}{L\epsilon}\,
    \Big[ a_m^{(Q)\dagger}\,\bm{C}_m^{(Q)}(1^-)\,\bm{Y}_q\,\bm{C}_n^{(q)}(1^-)\,a_n^{(q)}
    + a_m^{(q)\dagger}\,\bm{S}_m^{(q)}(1^-)\,\bm{Y}_q^\dagger\,\bm{S}_n^{(Q)}(1^-)\,a_n^{(Q)}
    \Big] \,, \\
   &= \frac{\sqrt 2\pi}{L\epsilon}\,a_m^{(Q)\dagger}\,\bm{C}_m^{(Q)}(1^-) \left(
    1 - \frac{v^2}{2M_{\rm KK}^2}\,\bm{\tilde Y}_q\,\bm{\tilde Y}_q^\dagger \right) \bm{Y}_q\,
    \bm{C}_n^{(q)}(1^-)\,a_n^{(q)} \,,    
\end{aligned}
\end{equation}
where in the second step we have used the mixed boundary conditions (\ref{naiveBCs}) at $t=1^-$. This result agrees with (\ref{g1mi}) by virtue of the definition of $\bm{\tilde Y}_q$ in (\ref{Xqdef}).

The Yukawa couplings derived in (\ref{g1mi}) will be used in the analysis in Section~\ref{sec:finitesum}. For the discussion in other parts of our paper, it will be necessary to keep the regulator $\eta$ non-zero until the sum over the tower of KK modes has been performed.

\section{Effective low-energy theory for $\bm{hgg}$ couplings}
\label{sec:Leff}

\begin{figure}
\begin{center}
\includegraphics[height=0.2\textwidth]{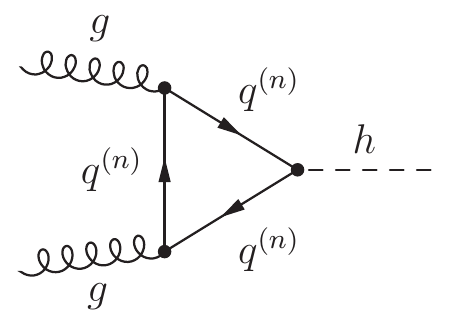}
\parbox{15.5cm}
{\caption{\label{fig:graph} 
Effective $hgg$ couplings induced by the exchange of virtual KK quarks.}}
\end{center}
\end{figure}

We are now ready to derive the effective low-energy Lagrangian for the Higgs-boson couplings to a pair of gluons, which are induced by the exchange of KK quarks. This Lagrangian is valid at energies below the scale $M_{\rm KK}$, at which these states can be integrated out. The relevant Feynman diagram arising at one-loop order is shown in Figure~\ref{fig:graph}. Since the gluon couplings to fermions are diagonal in the mass basis, a single quark state $q^{(n)}$ runs in the loop. Summing over the KK tower, we obtain 
\begin{equation}\label{Lhgg}
   {\cal L}_{hgg}^{\rm eff,KK} 
   = C_1^{\rm KK}(\mu)\,\frac{\alpha_s(\mu)}{12\pi}\,\frac{h}{v}\,G_{\mu\nu}^a\,G^{\mu\nu,a}
    - C_5^{\rm KK}(\mu)\,\frac{\alpha_s(\mu)}{8\pi}\,\frac{h}{v}\,
    G_{\mu\nu}^a\,\widetilde G^{\mu\nu,a} \,,
\end{equation}
where $\mu\le M_{\rm KK}$ is the scale at which the effective operators are renormalized, and $\widetilde G^{\mu\nu,a}=-\frac12\epsilon^{\mu\nu\alpha\beta}\,G_{\alpha\beta}^a$ (with $\epsilon^{0123}=-1$) is the dual field-strength tensor. Throughout this paper, $v$ denotes the value of the Higgs vacuum expectation value in the RS model, which differs from the SM value $v_{\rm SM}\approx 246$\,GeV by a small amount \cite{Bouchart:2009vq}. To first order in $v^2/M_{\rm KK}^2$, we obtain from the shift of the $W$-boson mass predicted by the RS model \cite{Casagrande:2008hr}
\begin{equation}\label{kappav}
   \kappa_v = \frac{v}{v_{\rm SM}}
   \approx 1 + \frac{m_W^2}{4 M_{\rm KK}^2} \left( L - 1 + \frac{1}{2L} \right) , 
\end{equation}
where $L\approx 34$. At the matching scale $\mu=M_{\rm KK}$, we find for the Wilson coefficients entering (\ref{Lhgg}) the one-loop results
\begin{equation}\label{matching}
\begin{aligned}
   C_1^{\rm KK}(M_{\rm KK}) 
   &= \sum_{q=u,d} {\sum_n}^\prime\,\frac{v\,\mbox{Re}(g_{nn}^q)}{m_{q_n}}\,A(\tau_{q_n}) \,, \\
    \qquad
   C_5^{\rm KK}(M_{\rm KK}) 
   &= \sum_{q=u,d} {\sum_n}^\prime\,\frac{v\,\mbox{Im}(g_{nn}^q)}{m_{q_n}}\,B(\tau_{q_n}) \,,
\end{aligned}
\end{equation}
where $\tau_{q_n}=4m_{q_n}^2/m_h^2-i\varepsilon$. The sums extend over all KK fermion states but exclude the SM fermions ($n\ne 1,2,3$). This is indicated by the prime on the sum symbol. Using the results from~\cite{Beneke:2002jn,Djouadi:2005gj}, we arrive at
\begin{equation}
   A(\tau) = \frac{3\tau}{2}\,\Big[ 1 + (1-\tau) \arctan^2\frac{1}{\sqrt{\tau-1}} \Big] \,,
    \qquad
   B(\tau) = \tau \arctan^2\frac{1}{\sqrt{\tau-1}} \,.
\end{equation}
Since the KK quarks are much heavier than the Higgs boson, it is sufficient to use the asymptotic values $A(\infty)=B(\infty)=1$ of these functions. It then follows that
\begin{equation}\label{C1C5}
   C_1^{\rm KK}(M_{\rm KK}) + i\,C_5^{\rm KK}(M_{\rm KK}) 
   = \sum_{q=u,d} {\sum_n}^\prime\,\,\frac{v g_{nn}^q}{m_{q_n}} \,.
\end{equation}
The real part of the sum determines $C_1^{\rm KK}$, while the imaginary part gives $C_5^{\rm KK}$.

We recall at this stage that each term in the sum (both the Yukawa couplings and the mass eigenvalues) depends on the regulator $\eta$ used to smear out the Higgs profile. In addition, one needs to worry about the convergence of the infinite sum over KK modes. Indeed, since the Yukawa couplings $g_{nn}^q$ are of ${\cal O}(1)$ and the masses of the KK modes are approximately evenly spaced multiples of the KK scale, naive dimensional analysis would suggest that the sum diverges logarithmically. In order to define it properly, one should therefore regularize the sum, for instance by introducing a cutoff on the highest KK level that is included. We thus define
\begin{equation}\label{Sigdef}
   \Sigma_q(N,\eta) = \sum_{n=1}^{3+6N} \frac{v g_{nn}^q}{m_{q_n}} \,,
\end{equation}
where the sum now includes the SM quarks ($n=1,2,3$) plus the first $N$ levels of KK modes. Each KK level contains six modes with ${\cal O}(v)$ mass splittings, while the different levels are split by an amount of ${\cal O}(M_{\rm KK})$. It will turn out that the sum over modes is finite despite naive expectation, because non-trivial cancellations happen among the six modes contained in each KK level. Nevertheless, the notion of a cutoff on the KK level will play an important role in our analysis.

The Wilson coefficients in (\ref{C1C5}) are now obtained as
\begin{equation}\label{doublelimit}
   C_1^{\rm KK}(M_{\rm KK}) + i\,C_5^{\rm KK}(M_{\rm KK}) 
   = \sum_{q=u,d} \bigg[ \lim_{N\to\infty,~\eta\to 0}\,\Sigma_q(N,\eta)
    - \sum_{n=1,2,3} \frac{v g_{nn}^q}{m_{q_n}} \bigg] \,,
\end{equation}
where the last term subtracts the contributions from the SM quarks, each of which equals 1 up to higher-order corrections in $v/M_{\rm KK}$. These contributions can be derived using the explicit expressions for the profile functions of SM quark fields near the IR brane given in \cite{Casagrande:2010si} as well as the relations for the $a_n^{(A)}$ vectors presented in \cite{Casagrande:2008hr}. We obtain
\begin{equation}\label{SMsums}
   \sum_{n=1,2,3} \frac{v g_{nn}^q}{m_{q_n}}
   = \mbox{Tr}\left( \frac{2\bm{X}_q}{\sinh2\bm{X}_q} \right) - \varepsilon_q \,,
\end{equation}
where the quantity 
\begin{equation}\label{vareps}
   \varepsilon_q = \mbox{Tr}\left( \bm{\delta}_Q + \bm{\delta}_q \right) 
    + {\cal O}\bigg( \frac{v^4}{M_{\rm KK}^4} \bigg)
   \approx \left(\delta_Q\right)_{33} + \left(\delta_q\right)_{33}
\end{equation}
contains some small corrections to the leading term. It receives its main contributions from the third-generation terms, because the quantities $\left(\delta_{Q,q}\right)_{nn}\propto m_{q_n}^2/M_{\rm KK}^2$ defined in \cite{Casagrande:2008hr} are chirally suppressed for all light quarks. To a very good approximation
\begin{equation}
   \left( \delta_U \right)_{33} 
   = \frac{m_t^2}{M_{\rm KK}^2} \left[ \frac{1}{F^2(c_{u_3})} \sum_{i=1}^3\,
    \frac{1}{1-2c_{u_i}}\,\frac{|(Y_u)_{3i}|^2}{|(Y_u)_{33}|^2}
    - \frac{1}{1-2c_{u_3}} \left( 1 - \frac{F^2(c_{u_3})}{3+2c_{u_3}} \right) \right] , 
\end{equation}
and $\left( \delta_u \right)_{33}$ is given by the same formula with $c_{u_i}\to c_{Q_i}$ and $(Y_u)_{3i}\to(Y_u)_{i3}$. Analogous expressions hold in the down-type quark sector. The function
\begin{equation}
   F^2(c) = \frac{1+2c}{1-\epsilon^{1+2c}}
\end{equation}
determines the values of the fermion profiles on the IR brane \cite{Grossman:1999ra,Gherghetta:2000qt}. Note that in the approximation used here both $\varepsilon_q$ and the zero-mode sum in (\ref{SMsums}) are real and do not contribute to the Wilson coefficient $C_5^{\rm KK}$ in (\ref{doublelimit}).   

The cross section for Higgs-boson production in gluon-gluon fusion can now be written as
\begin{equation}\label{sigmaRS}
    \sigma(gg\to h)_{\rm RS} 
    = \frac{\left| \kappa_g \right|^2 + \left| \kappa_{5g} \right|^2}{\kappa_v^2}\, 
     \sigma(gg\to h)_{\rm SM} \,,
\end{equation}
where $\kappa_v$ has been defined in (\ref{kappav}), while
\begin{equation}\label{kappas}
\begin{aligned}
   \kappa_g 
   &= \frac{C_1^{\rm KK}(m_h)+\sum_{i=t,b}\,\mbox{Re}(\kappa_i)\,A(\tau_i)}%
           {\sum_{i=t,b}\,A(\tau_i)} \,, \\
   \kappa_{5g} 
   &= \frac32\,\frac{C_5^{\rm KK}(m_h)+\sum_{i=t,b}\,\mbox{Im}(\kappa_i)\,B(\tau_i)}%
                    {\sum_{i=t,b}\,A(\tau_i)} \,.
\end{aligned}
\end{equation}
Here $\kappa_t=v g_{33}^u/m_t$ and $\kappa_b=v g_{33}^d/m_b$ encode the modifications of the Higgs-boson couplings to top and bottom quarks with respect to their SM values. The former quantity is given to high accuracy by \cite{Goertz:2011hj}
\begin{equation}\label{kappat}
   \kappa_t\approx 1 - \frac{v^2}{3M_{\rm KK}^2}\,
    \frac{\left(\bm{Y}_u\bm{Y}_u^\dagger\bm{Y}_u\right)_{33}}{\left(Y_u\right)_{33}} 
    - \left(\delta_U\right)_{33} - \left(\delta_u\right)_{33} \,,
\end{equation}
and an analogous expression with $u\to d$ and $U\to D$ holds for $\kappa_b$.

The Wilson coefficients $C_{1,5}^{\rm KK}$ must be evolved from the high KK scale to the Higgs-boson mass scale. Their scale dependence is governed by evolution equations. If the KK contributions are dominated by the corrections due to the lowest-lying modes, it is practical to integrate out all KK modes at a common scale of order $M_{\rm KK}$, as already done in (\ref{matching}) above. The evolution from this scale down to the weak scale is described by the renormalization-group equations of the SM. For the QCD evolution of the Wilson coefficient $C_1^{\rm KK}(\mu)$, one obtains the exact result \cite{Inami:1982xt,Ahrens:2008nc}
\begin{equation}
   \frac{C_1^{\rm KK}(\mu)}{C_1^{\rm KK}(M_{\rm KK})} 
   = \frac{\beta\big(\alpha_s(\mu)\big)/\alpha_s^2(\mu)}%
          {\beta\big(\alpha_s(M_{\rm KK})\big)/\alpha_s^2(M_{\rm KK})}
   = 1 + \frac{13}{14}\,\frac{\alpha_s(\mu)-\alpha_s(M_{\rm KK})}{\pi} + \dots \,.
\end{equation}
In practice, the evolution from the KK scale of several TeV down to $\mu\approx m_h$ has only a small effect of about 1\% on the value of $C_1^{\rm KK}$, since no leading logarithms appear in this result. Since the operator multiplying the coefficient $C_5^{\rm KK}$ in (\ref{Lhgg}) is connected by the Adler-Bell-Jackiw anomaly to current operators with vanishing QCD anomalous dimension, it follows that this coefficient is scale independent in QCD, i.e., $C_5^{\rm KK}(\mu)=C_5^{\rm KK}(M_{\rm KK})$.

Much of the analysis in the present paper is concerned with the fact that the order in which the limits $\eta\to 0$ and $N\to\infty$ in (\ref{doublelimit}) are taken is not irrelevant, because these two limits do not commute. In our previous work \cite{Casagrande:2010si}, we have evaluated the Yukawa couplings in (\ref{gdef}) mode by mode, taking the regulator $\eta$ to zero after computing the relevant overlap integrals. We have then numerically evaluated the contributions of the first few KK levels to the Higgs-boson production cross section, observing that the sum over modes converges and approaches a limiting value after the summation over several KK levels. In essence, this approach corresponds to taking the limit $\eta\to 0$ first, and hence
\begin{equation}\label{ourS}
   \Sigma_q^{\rm (CGHNP)} = \lim_{N\to\infty} \Big[ \lim_{\eta\to 0}\,\Sigma_q(N,\eta) \Big]
   = \lim_{N\to\infty}\,\sum_{n=1}^{3+6N}\,\frac{v g_{nn}^q}{m_{q_n}} \bigg|_{\eta\to 0} \,.
\end{equation}
Yet we did not attempt to derive an analytical expression for the infinite sum. In the approach taken in \cite{Azatov:2010pf}, on the other hand, one considers the infinite sum over modes from the very beginning. This is accomplished by means of completeness relations for the fermion profiles. The regularization of the Higgs profile is taken to zero at the end of the calculation. It follows that in this approach
\begin{equation}\label{theirS}
   \Sigma_q^{\rm (ATZ)} = \lim_{\eta\to 0} \left[ \lim_{N\to\infty} \Sigma_q(N,\eta) \right]
   = \lim_{\eta\to 0}\,\sum_{n=1}^\infty\,\frac{v g_{nn}^q}{m_{q_n}} \,.
\end{equation}
In the following two sections, we will derive closed analytical expressions for the limiting values in (\ref{ourS}) and (\ref{theirS}). In Section~\ref{sec:interp}, we will then discuss the physical interpretation of these results.

\section{Summing first over the infinite KK tower}
\label{sec:Azatov}

We begin with the study of the sum in (\ref{theirS}), in which the regulator on the Higgs profile is removed after the infinite sum over KK modes has been performed. The authors of \cite{Azatov:2010pf} succeeded to evaluate this sum using the completeness relations ($A=L,R$)
\begin{equation}\label{completeness}
   \sum_n\,{\cal Q}_A^{(n)}(t)\,{\cal Q}_A^{(n)\dagger}(t') = \delta(t-t')
\end{equation}
for the fermion profiles. The analysis relied on a perturbative treatment of the Yukawa couplings, which yields the first non-trivial contribution in the expansion in powers of $v/M_{\rm KK}$. A subtlety in this calculation, worth recalling, is that in the perturbative approach the odd fermion profiles $\bm{S}_n^{(A)}(t)$ vanish on the IR brane. As a result, the contribution of each individual mode, and hence of any truncated KK sum, vanishes in the limit where the width of the Higgs-boson profile is taken to zero. A non-zero result for the ${\cal O}(v^2/M_{\rm KK}^2)$ correction to (\ref{theirS}) is obtained only after summing over the infinite tower of KK states. In our notation, the main result of \cite{Azatov:2010pf} reads  
\begin{equation}\label{eq28}
   \Sigma_q^{\rm (ATZ)} 
   = \mbox{Tr}\left( 1 + \frac{\bm{X}_q^2}{3} + \dots \right) . 
\end{equation}

An elegant way to derive a closed expression for the sum (\ref{theirS}), valid to all orders in $v/M_{\rm KK}$, is to relate it to the 5D fermion propagator of the theory. We study propagator functions built using the 6-component spinors
\begin{equation}
   {\cal Q}_A(t,x) = \sum_n\,{\cal Q}_A^{(n)}(t)\,q_A^{(n)}(x) \,,
\end{equation}
where the profile functions ${\cal Q}_A^{(n)}(t)$ have been defined in (\ref{notat}). The grand, $6\times 6$ propagator in the mixed momentum/position representation \cite{Puchwein:2003jq,Carena:2004zn} is
\begin{eqnarray}
\begin{aligned}
   i\bm{S}^q(t,t';p) &= \int d^4x\,e^{ip\cdot x}\,\langle\,0|\,{\bm T}
    \big( \,{\cal Q}_L(t,x) + {\cal Q}_R(t,x) \big) 
    \big( \,\bar{\cal Q}_L(t',0) + \bar{\cal Q}_R(t',0) \big)\,|0\,\rangle \\
   &\hspace{-15mm}= \sum_n \left[ \,{\cal Q}_L^{(n)}(t)\,\frac{1-\gamma_5}{2} 
    + {\cal Q}_R^{(n)}(t)\,\frac{1+\gamma_5}{2} \right] \frac{i}{\rlap{\hspace{0.3mm}/}{p}-m_{q_n}}
    \left[ \,{\cal Q}_L^{(n)\dagger}(t')\,\frac{1+\gamma_5}{2} 
    + {\cal Q}_R^{(n)\dagger}(t')\,\frac{1-\gamma_5}{2} \right] , \hspace{4mm}
\end{aligned}
\end{eqnarray}
where ${\bm T}$ denotes time ordering. In our notation, the Dirac operator takes the form
\begin{equation}
   {\cal D} = \rlap{\hspace{0.3mm}/}{p} - M_{\rm KK}\,\gamma_5\,\frac{\partial}{\partial t}
    - M_{\rm KK}\,{\cal M}_q(t) \,,
\end{equation}
with the generalized mass matrix defined in (\ref{Mqdef}). Using the bulk EOMs (\ref{eoms}) and the completeness relations (\ref{completeness}), it is straightforward to show that
\begin{equation}\label{Dirac}
   {\cal D}\,\bm{S}^q(t,t';p) = \delta(t-t') \,.
\end{equation}

Since with our exact treatment of Yukawa interactions there are no massless zero modes of the fermion fields, the propagator does not exhibit a singularity at $p^2=0$. We can therefore study the special limit $p^\mu\to 0$ without complications. We obtain
\begin{equation}\label{S0prop}
   \bm{S}^q(t,t';0) = - \left[ \bm{\Delta}_{RL}^q(t,t')\,\frac{1+\gamma_5}{2}
    + \bm{\Delta}_{LR}^q(t,t')\,\frac{1-\gamma_5}{2} \right] ,
\end{equation}
where
\begin{equation}
   \bm{\Delta}_{RL}^q(t,t') 
   = \sum_n\,\frac{1}{m_{q_n}}\,{\cal Q}_R^{(n)}(t)\,{\cal Q}_L^{\dagger (n)}(t') \,, \qquad
   \bm{\Delta}_{LR}^q(t,t') = \bm{\Delta}_{RL}^{q\,\dagger}(t',t) \,.
\end{equation}
It now follows from the second relation in (\ref{gdef}) that the infinite sum in (\ref{theirS}), for fixed non-zero $\eta$, can be expressed as \cite{Falkowski:2007hz}
\begin{equation}\label{gsum}
   \lim_{N\to\infty} \Sigma_q(N,\eta) 
   = \int_\epsilon^1\!dt\,\delta^\eta(t-1)\,T_{RL}^q(t,t) \,,
\end{equation}
where
\begin{equation}\label{TRdef}
   T_{RL}^q(t,t') = \frac{v}{\sqrt 2}\,\mbox{Tr}\left[
    \bigg( \begin{array}{cc} \bm{0} & \bm{Y}_q \\ 
           \bm{Y}_q^\dagger & \bm{0} \end{array} \bigg)\,\bm{\Delta}_{RL}^q(t,t') \right] .
\end{equation}
Notice that only the off-diagonal blocks of the propagator enter in this result. 

The Dirac equation (\ref{Dirac}) for the 5D propagator implies the differential equation
\begin{equation}\label{Delta_eom}
   \bigg[ \frac{\partial}{\partial t} + {\cal M}_q(t) \bigg]\,\bm{\Delta}_{RL}^q(t,t') 
   = \frac{1}{M_{\rm KK}}\,\delta(t-t') \,.
\end{equation}
We will construct solutions to this equation assuming first that $t\ne t'$, distinguishing the cases where $t>t'$ and $t<t'$. Later, these solutions are patched together by means of the jump condition
\begin{equation}\label{jump}
   \lim_{\delta\to 0}\,\Big[ 
   \bm{\Delta}_{RL}^q(t+\delta,t) - \bm{\Delta}_{RL}^q(t-\delta,t) \Big]
   = \frac{1}{M_{\rm KK}} \,.
\end{equation}
Since we have regularized the Higgs profile, the boundary conditions on the UV and IR branes take the simple form
\begin{equation}\label{BCs}
   ( \bm{1} \quad \bm{0} )\,\bm{\Delta}_{RL}^q(\epsilon,t') 
   = ( \bm{1} \quad \bm{0} )\,\bm{\Delta}_{RL}^q(1,t') = ( \bm{0} \quad \bm{0} ) \,,
\end{equation}
which follows since the $Z_2$-odd profile functions $\bm{S}_n^{(A)}(t)$, which sit in the upper components of the right-handed spinors in (\ref{notat}), vanish on the branes. The three equations above determine the matrix function $\bm{\Delta}_{RL}^q(t,t')$ completely.

For our purposes, it suffices to construct the solution in the region close to the IR brane, where $1-\eta\le t,t'\le 1$. For $t>t'$, equation (\ref{Delta_eom}) is solved by the $t$-ordered exponential
\begin{equation}
   \bm{\Delta}_{RL}^q(t,t') \big|_{t>t'}
   = \bm{T} \exp\left[ - \int_1^t\!ds\,{\cal M}_q(s) \right]\,\bm{\Delta}_{RL}^q(1,t') \,,
\end{equation}
where (\ref{BCs}) implies that the boundary function $\bm{\Delta}_{RL}^q(1,t')$ has only lower components. The ordering symbol is required since the matrices
\begin{equation}\label{intM}
   \int_1^t\!ds\,{\cal M}_q(s) 
   = \ln t\,\bigg( \begin{array}{cc} \bm{c}_Q & \bm{0} \\ 
                  \bm{0} & -\bm{c}_q \end{array} \bigg) 
    - \frac{v}{\sqrt 2 M_{\rm KK}}\,\bar\theta^\eta(t-1)\,
    \bigg( \begin{array}{cc} \bm{0} & \bm{Y}_q \\ 
           \bm{Y}_q^\dagger & \bm{0} \end{array} \bigg) \,,
\end{equation}
with $\bar\theta^\eta(t-1)$ as defined in (\ref{thetadef}), do not commute at different $t$ values. For coordinates $t$ near the IR brane, however, the first term in (\ref{intM}) can be neglected, and we can approximate 
\begin{equation}\label{sol1}
   \bm{\Delta}_{RL}^q(t,t') \big|_{t>t'}
   = \exp\left[ \frac{v}{\sqrt 2 M_{\rm KK}}\,\bar\theta^\eta(t-1)\,
    \bigg( \begin{array}{cc} \bm{0} & \bm{Y}_q \\ 
           \bm{Y}_q^\dagger & \bm{0} \end{array} \bigg) 
    + {\cal O}(\eta) \right] \bm{\Delta}_{RL}^q(1,t') \,,
\end{equation}
where the harmless ${\cal O}(\eta)$ terms can be dropped. The resulting matrix exponential can then be evaluated in terms of hyperbolic trigonometric functions. We then lower the value of $t$, implement the jump condition (\ref{jump}) when $t$ crosses $t'$, and finally derive an expression for $\bm{\Delta}_{RL}^q(1-\eta,t')$, which still contains two unknown coefficients that are matrices in generation space and functions of $t'$. They can be determined using the boundary condition on the UV brane. For $t<1-\eta$ the generalized mass matrix ${\cal M}_q(t)$ in (\ref{Mqdef}) is diagonal, since the contribution from the Higgs profile vanishes. The first boundary condition in (\ref{BCs}) therefore implies that $(\bm{1}\quad\bm{0})\,\bm{\Delta}_{RL}^q(1-\eta,t')=(\bm{0}\quad\bm{0})$ as long as $t'\ge 1-\eta$, and this equation yields the desired constraints on the propagator function. 

In (\ref{TRdef}) we only need the off-diagonal blocks of the function $\bm{\Delta}_{RL}^q(t,t')$ for $1-\eta<t,t'<1$. We find 
\begin{equation}\label{beauty}
   T_{RL}^q(t,t') 
   = \mbox{Tr}\left[ \frac{\bm{X}_q}{\sinh\bm{X}_q}\,
    \cosh\left(\bm{X}_q \left( 1 - \int_{t_<}^{t_>}\!ds\,\delta^\eta(s-1) \right)\!\right)
    \right] ,
\end{equation}
where $t_<=\mbox{min}(t,t')$ and $t_>=\mbox{max}(t,t')$. Notice that, for the special case where $t=t'$, the result in expression (\ref{beauty}) is independent of both $t$ and $\eta$. This means that in (\ref{gsum}) only the normalization of the Higgs profile enters, which is independent of its shape and equal to 1. We conclude that the sum in (\ref{gsum}) is in fact independent of $\eta$, and hence 
\begin{equation}\label{SATZ}
   \Sigma_q^{\rm (ATZ)}
   = \mbox{Tr}\left( \bm{X}_q\coth\bm{X}_q \right) 
   = \mbox{Tr}\left( 1 + \frac{\bm{X}_q^2}{3} - \frac{\bm{X}_q^4}{45} \pm \dots \right) .
\end{equation}
The first non-trivial term in the Taylor expansion agrees with (\ref{eq28}). Note that this result is real, and hence it only contributes to the Wilson coefficient $C_1^{\rm KK}$. Remarkably, the infinite sum of KK states converges despite of the fact that it is superficially divergent. This hints to the existence of intricate cancellations between different contributions to the sum. From (\ref{doublelimit}) and (\ref{SMsums}), we now obtain
\begin{equation}\label{funny}
\begin{aligned}
   C_1^{\rm KK}(M_{\rm KK})
   &= \sum_{q=u,d}\,\Big[ \mbox{Tr}\left( \bm{X}_q\tanh\bm{X}_q \right)
    + \mbox{Re}\,\varepsilon_q \Big] \\
   &= \sum_{q=u,d}\,\left[ \mbox{Tr}\left( \bm{X}_q^2 - \frac{\bm{X}_q^4}{3} 
    + \dots \right) + \mbox{Re}\,\varepsilon_q \right] , 
\end{aligned}
\end{equation}
and 
\begin{equation}\label{C5res}
   C_5^{\rm KK}(M_{\rm KK}) = \mbox{Im}\left(\varepsilon_u + \varepsilon_d\right) .
\end{equation}
Recall that the correction terms $\varepsilon_q$ are very small, and to an excellent approximation they  are real, see (\ref{vareps}). The Wilson coefficient $C_5^{\rm KK}$ is therefore predicted to be tiny, of order $v^4/M_{\rm KK}^4$ at most. The contribution to the coefficient $C_1^{\rm KK}$ is positive and hence yields to an enhancement of the $gg\to h$ amplitude compared with the SM.

\section{Removing first the regulator on the Higgs profile}
\label{sec:finitesum}

We now consider the sum in (\ref{ourS}), for which compared to (\ref{theirS}) the limits $N\to\infty$ and $\eta\to 0$ are taken in the opposite order. In this case one first considers a finite sum over KK modes and derives the relevant Yukawa coupling for each mode by regularizing the Higgs profile, computing the overlap integral in (\ref{gdef}), and then taking the limit $\eta\to 0$. The relevant Yukawa couplings in that limit have been given in (\ref{g1mi}).

Because the sum in (\ref{ourS}) extends over a finite number of KK levels, it is not possible to use the elegant method of 5D propagators described in the previous section. In order to obtain a closed expression nevertheless, we adopt the strategy of first finding a solution in a special case, where the bulk EOMs and the eigenvalue equation determining the masses of the KK modes can be solved analytically. We will then argue that, as in the case (\ref{SATZ}), the solution is independent of the bulk mass parameters. 

The special case we consider is that of one generation of fermion fields, whose bulk mass parameters vanish: $c_Q=c_u=c_d=0$. The EOMs for the profile functions are then solved in terms of simple trigonometric functions, which can be evaluated in the limit $\epsilon\to 0$ (recall that $\epsilon\approx 10^{-15}$ is  tiny in the RS model). We find that the KK masses in units of $M_{\rm KK}$ are given by the solutions to the eigenvalue equation
\begin{equation}
   \tan^2\!x_n = \tanh^2\!X_q \,; \qquad
   X_q = \frac{v}{\sqrt2 M_{\rm KK}}\,Y_q \,,
\end{equation}
where without loss of generality we assume that $Y_q$ is real and positive. This equation can be solved to give
\begin{equation}\label{xn}
   x_n = \left\{ \begin{array}{rl} 
    \displaystyle\frac{n-1}{2}\,\pi + x_1 \,; & \quad n=1,3,5,\dots, \\[4mm]
    \displaystyle\frac{n}{2}\,\pi - x_1 \,; & \quad n=2,4,6,\dots, 
   \end{array} \right.
\end{equation}
where $x_1=\arctan(\tanh X_q)$ denotes the mass of the zero mode (the ``SM quark'') in units of the KK scale. The corresponding even profile functions are
\begin{equation}
   \sqrt{\frac{2\pi}{L\epsilon}}\,C_n^{(Q)}(t)\,a_n^{(Q)} = \cos(x_n t) \,, \qquad
   \sqrt{\frac{2\pi}{L\epsilon}}\,C_n^{(q)}(t)\,a_n^{(q)} = \pm \cos(x_n t) \,,
\end{equation}
where the upper (lower) signs hold for odd (even) values of $n$. Inserting these results into (\ref{g1mi}) and using (\ref{xn}), it follows that
\begin{equation}
   g_{nn}^q 
   = \frac{\pm 1}{\sqrt 2}\,\frac{Y_q}{\cosh^2\!X_q}\,\cos^2\!x_n 
   = \frac{\pm 1}{\sqrt 2}\,\frac{Y_q}{\cosh 2X_q} \,.
\end{equation}
We can now readily compute the sum over KK modes required in (\ref{ourS}), with the result that
\begin{equation}\label{toysum}
   \lim_{\eta\to 0} \Sigma_q(N,\eta) \Big|_{c_A=0}^{\rm 1~gener.}
   = \sum_{n=1}^{1+2N}\,\frac{v g_{nn}^q}{m_{q_n}} \bigg|_{\eta\to 0}
   = \frac{X_q}{\cosh 2X_q} \left[ \frac{1}{x_1}
    + \sum_{k=1}^N \left( \frac{1}{k\pi+x_1} - \frac{1}{k\pi-x_1} \right) \right] .
\end{equation}
In the one-generation case, $n=1$ refers to the zero mode, while $n\ge 2$ labels the KK modes. The first term in the bracket of the final expression arises from the ``SM quark'', while the sum is over pairs of KK modes belonging to the $k^{\rm th}$ KK level. Note that for large $k$ the individual terms in the sum fall off only like $1/k$, but each pair combines to a contribution decreasing like $1/k^2$. Hence, as long as we sum over complete levels of KK states first, the sum over $k$ is convergent, and it is possible to take the limit $N\to\infty$. We thus obtain for (\ref{ourS}) in the case of a single fermion generation with vanishing bulk masses
\begin{equation}\label{magic}
   \Sigma_q^{\rm (CGHNP)} \Big|_{c_A=0}^{\rm 1~gener.}
   = \frac{X_q\coth X_q}{\cosh 2X_q} \,,
\end{equation}
which differs from the finite sum in (\ref{toysum}) by terms of ${\cal O}(1/N)$.

For one generation, it is not difficult to solve the EOMs numerically also for general $c_A\ne 0$. The profile functions and eigenvalue equation in this case involve Bessel functions. We have confirmed in this way that the simple answer (\ref{magic}) still holds in this more general case. Given the similarity of the above result with equation (\ref{SATZ}) derived in the previous section, we conjecture that in the case of three generations
\begin{equation}\label{conjecture}
   \Sigma_q^{\rm (CGHNP)} 
   = \mbox{Tr}\left( \frac{\bm{X}_q\coth\bm{X}_q}{\cosh 2\bm{X}_q} \right)
   = \mbox{Tr}\left( 1 - \frac{5\bm{X}_q^2}{3} + \frac{119\bm{X}_q^4}{45} \mp \dots \right) .
\end{equation}
Once again the answer is real, such that only $C_1^{\rm KK}$ receives a contribution. Combining (\ref{doublelimit}), (\ref{SMsums}), and (\ref{conjecture}), we then find
\begin{equation}\label{central_result}
\begin{aligned}
   C_1^{\rm KK}(M_{\rm KK}) 
   &= \sum_{q=u,d} \left[ - \mbox{Tr}\left( \frac{\bm{X}_q\tanh\bm{X}_q}{\cosh 2\bm{X}_q} \right)
    + \mbox{Re}\,\varepsilon_q \right] \\
   &= \sum_{q=u,d} \left[ \mbox{Tr}\left( - \bm{X}_q^2 + \frac{7\bm{X}_q^4}{3} + \dots \right)
    + \mbox{Re}\,\varepsilon_q \right] ,
\end{aligned}
\end{equation}
while $C_5^{\rm KK}$ is still given by (\ref{C5res}). This time the contribution to $C_1^{\rm KK}$ is negative and thus yields a suppression of the $gg\to h$ amplitude compared with the SM. We have checked that this formula indeed reproduces our numerical results obtained in \cite{Casagrande:2010si}. Incidentally, the first term in the expansion in powers of $\bm{X}_q^2$ has the opposite sign from the result (\ref{funny}), which was first obtained in \cite{Azatov:2010pf}. 

The conjecture that equation (\ref{magic}) can be generalized to the case of three generations will be proved elsewhere. We note, however, that symmetry arguments help us to constrain the form of the answer. It is plausible to assume that, after the limit $N\to\infty$ has been taken, the result for the sum (\ref{ourS}) should depend only on the fundamental parameters of the underlying 5D theory, i.e., the bulk mass parameters and Yukawa couplings. All reference to the properties of individual states, such as their masses or profile functions, should disappear. We express this assertion by writing $\Sigma_q^{\rm (CGHNP)}=\Sigma(\bm{Y}_q,\bm{c}_Q,\bm{c}_q)$. The dependence of this quantity on the fundamental parameters can be constrained using symmetry arguments. The EOMs (\ref{eoms}) and boundary conditions (\ref{IRbcs}) are valid in an arbitrary basis, in which the bulk mass matrices $\bm{c}_A$ are not necessarily diagonal. These relations are invariant under a set of three global symmetries. The first one is a symmetry under the exchange of $SU(2)$ doublets and singlets along with $\bm{Y}_q\leftrightarrow\bm{Y}_q^\dagger$. In addition, there are two symmetries related to unitary transformations of the Yukawa and bulk mass matrices. When combined with the fact that in the one-generation case the result is found to be independent of the bulk mass parameters, these symmetries imply that the quadratic term in $\bm{X}_q$ must indeed be of the simple form shown in (\ref{conjecture}). Using symmetry considerations alone, we can however not exclude a dependence on the parameters $\bm{c}_A$ starting at ${\cal O}(\bm{X}_q^4)$, provided that it cancels in the case of one generation.

\section{Reconciling the results}
\label{sec:interp}

The discussion of the previous two sections shows that the calculations presented in \cite{Casagrande:2010si} and~\cite{Azatov:2010pf} are free of mistakes. The results for the effective $hgg$ couplings derived in these two papers differ, in magnitude and sign, because both groups considered different orders in which the limits $\eta\to 0$ and $N\to\infty$ in (\ref{doublelimit}) were taken, and the two limits do not commute. Hence, the question poses itself which of the orders of limits is the more reasonable one from a physical point of view. We will address this question in Section~\ref{sec:cutoff}.

A second, equally puzzling question arises once we realize that the non-commutativity of the limits $\eta\to 0$ and $N\to\infty$ implies that the origin of the discrepancy between the results~(\ref{SATZ}) and (\ref{conjecture}) must be due to contributions from KK modes with large $n$ and hence very heavy masses, and related to the behavior of the profile functions for such modes very close to the IR brane. The fact that in both approaches one finds that the infinite sum over KK modes converges, signaling that heavy modes decouple, raises the question: how is it possible that modes with very heavy masses give an ${\cal O}(1)$ contribution to the sum, which is large enough to change the sign of the answer?

In order to understand the physics behind this effect, it is useful to study a toy model that can be solved exactly also for finite $\eta$. To this end, we consider again the case of a single fermion generation and vanishing bulk mass parameters, which proved so useful for our analysis in Section~\ref{sec:finitesum}. To obtain analytic solutions to the bulk EOMs for the fermion profiles we adopt a particularly simple form for the regularized Higgs profile, namely a box of width $\eta$:
\begin{equation}\label{Hbox}
   \delta^\eta(t-1) = \left\{ \, \begin{array}{cl} \displaystyle
    \frac{1}{\eta} \,\,; & \quad \mbox{for~~} 1-\eta < t < 1 \,, \\[4mm]
    0 \,\,; & \quad \mbox{for~~} \epsilon\le t < 1-\eta \,.
    \end{array} \right. 
\end{equation}
Moreover, since for $c_A=0$ the wave functions are non-singular near the UV brane, it is possible to set $\epsilon=0$. In Appendix~\ref{sec:toy} we derive the explicit forms of the bulk profiles for this toy model and the eigenvalue equation that determines the masses of the KK modes. We then present a formula for the sum $\Sigma_q(N,\eta)$ entering (\ref{doublelimit}). It turns out that the nature of the solution differs depending on whether $x_n<z$ or $x_n>z$, where we have introduced the abbreviation $z=v Y_q/(\sqrt2 M_{\rm KK}\eta)=X_q/\eta$. The latter condition corresponds to masses $m_{q_n}>v Y_q/(\sqrt2 \eta)=M_{\rm weak}/\eta$, where $M_{\rm weak}\equiv v Y_q/\sqrt2$ is of order the weak scale, since we as usual assume $Y_q={\cal O}(1)$ in the anarchic RS model (recall that we can choose $Y_q$ real and positive). The appearance of the scale $M_{\rm weak}/\eta$, which for very small $\eta$ lies far above the TeV scale, will be of crucial importance to the resolution of the puzzle.

We find that the mass eigenvalues in our toy model are determined by the conditions
\begin{equation}\label{eigenvals}
   \tan\big[x_n(1-\eta)\big] 
   = \left\{ \begin{array}{ll}
    \displaystyle 
    \pm\sqrt{\frac{z\mp x_n}{z\pm x_n}} \tanh\big[\sqrt{z^2-x_n^2}\,\eta\big] \,; 
    & \quad \mbox{for~~} x_n<z \,, \\[5mm]
    \displaystyle 
    - \sqrt{\frac{x_n\mp z}{x_n\pm z}} \tan\big[\sqrt{x_n^2-z^2}\,\eta\big] \,;
    & \quad \mbox{for~~} x_n>z \,,
   \end{array} \right.
\end{equation}
where as before the upper (lower) signs hold for odd (even) values of $n$. These equations can easily be solved numerically. The sum over the KK contributions to the effective $hgg$ couplings is then obtained by evaluating the sum 
\begin{equation}\label{toysum2}
\begin{aligned}
   \Sigma_q(N,\eta) 
   &= \sum_{n=1}^{1+2N}\,\frac{v g_{nn}^q}{m_{q_n}}
   = \sum_{n=1}^{1+2N} \left[ 1 - N_n^2 \left( 1\mp \frac{\eta z}{x_n}\cos\big[2x_n(1-\eta)\big]
    \right) \right] , \\
   N_n^{-2} &= 1 + \frac{z}{z^2-x_n^2} \left[ \pm\frac12 \sin\big[2x_n(1-\eta)\big]
    - \eta\left( z\mp x_n\cos\big[2x_n(1-\eta)\big] \right) \right] .
\end{aligned}
\end{equation}

\begin{figure}
\begin{center}
\includegraphics[height=0.45\textwidth]{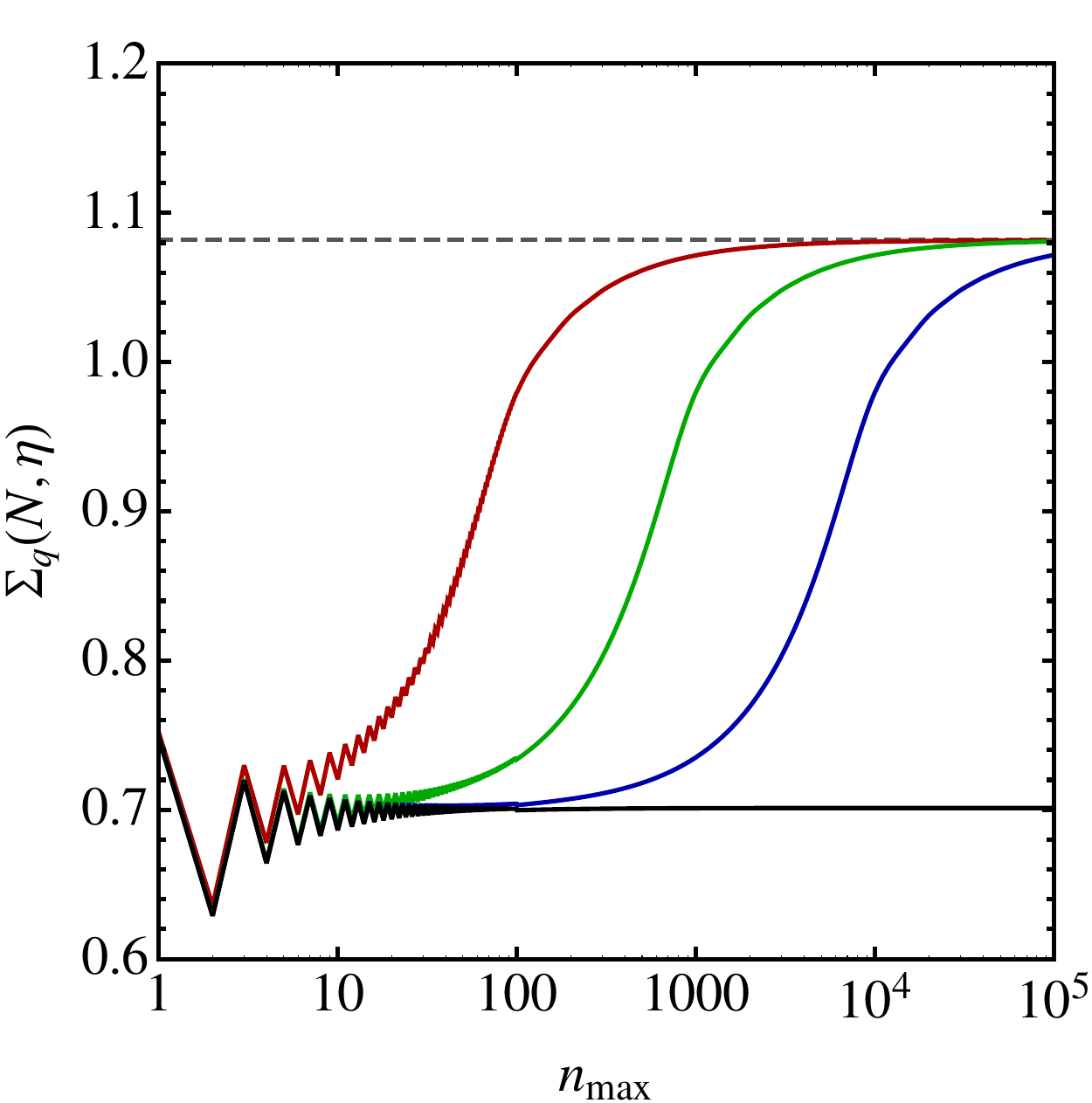} 
\quad
\includegraphics[height=0.453\textwidth]{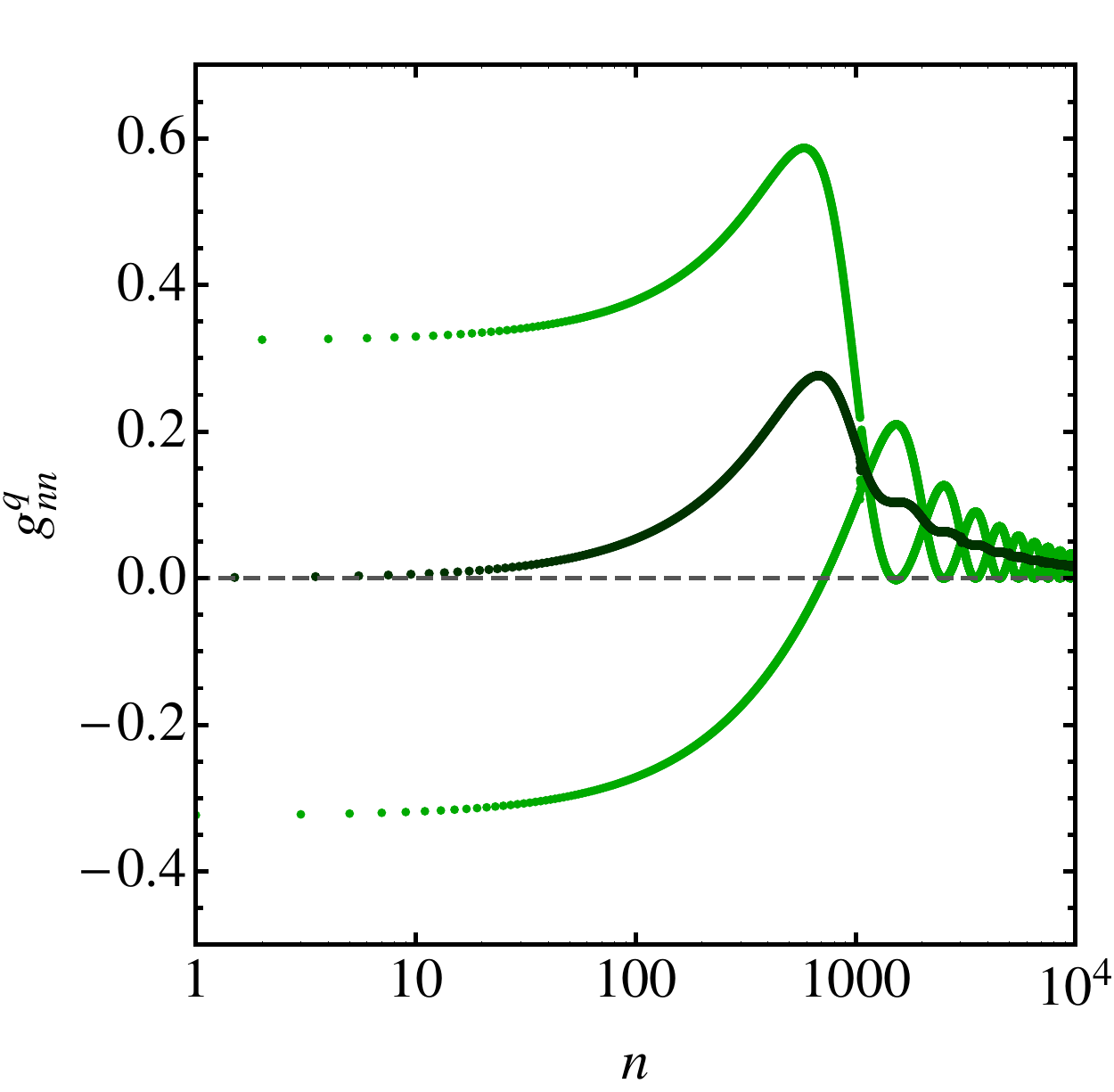} 
\parbox{15.5cm}
{\caption{\label{fig:surprise} 
{\em Left:} 
Partial sums $\Sigma_q(N,\eta)$ for different values of $n_{\rm max}=1+2N$ and $\eta$ in the toy model with one generation and vanishing bulk mass parameters. The curves refer to $\eta=10^{-2}$ (red), $10^{-3}$ (green), $10^{-4}$ (blue), and 0 (black). 
{\em Right:} 
Yukawa couplings $g_{nn}^q$ of the KK fermion states in the toy model for $\eta=10^{-3}$. The dark points show the Yukawa couplings averaged over the pair of modes in each KK level.}}
\end{center}
\end{figure}

In the left plot in Figure~\ref{fig:surprise}, we show numerical results for the sum (\ref{toysum2}) as a function of the number of KK states included, for four different values of $\eta$. For the purpose of illustration we take $X_q=0.5$. We observe that for low values of $N$ the sum quickly converges toward a value close to the result (\ref{ourS}), which equals 0.701 in the present case. But then there is an intermediate region, roughly given by the range $0.1/\eta<N<10/\eta$, in which the value of the sum changes by an ${\cal O}(1)$ amount. After this transition region, the sum converges to the value corresponding to the result (\ref{SATZ}), which equals 1.082 for our choice of $X_q$. In order to understand the origin of the three regions -- in particular the appearance of the intermediate region, in which the sum grows by an ${\cal O}(1)$ amount despite of the fact that the corresponding KK masses are extremely heavy -- we show in the right plot the values of the Yukawa couplings $g_{nn}^q$ in our toy model (for the case where $\eta=10^{-3}$). For low values of $n$, the couplings for each pair of KK states in the same KK level have equal values and opposite signs. In this case, like in (\ref{toysum}), the contribution from the $k^{\rm th}$ KK level is 
\begin{equation}
   \frac{X_q}{\cosh 2X_q} \left( \frac{1}{k\pi+x_1} - \frac{1}{k\pi-x_1} \right) 
   = - \frac{X_q}{\cosh 2X_q}\,\frac{2x_1}{k^2\pi^2-x_1^2} \,.
\end{equation}
In this region the sum over $k$ is convergent and dominated by the contributions from the first few KK levels. The situation changes drastically in the intermediate region, where the average coupling $g_{\rm avg}(k)$ in each KK level, shown by the dark points in the right plot, no longer vanishes. There is thus a range of logarithmic growth of the sum, over which its contribution can be estimated as 
\begin{equation}\label{loggrowth}
   \sum_{k=N_1}^{N_2}\,\frac{g_{\rm avg}(k)}{k\pi} 
   \approx \frac{\langle g_{\rm avg}\rangle}{\pi}\,\ln\frac{N_2}{N_1} \,,
\end{equation}
where $\langle g_{\rm avg}\rangle$ denotes the mean value of $g_{\rm avg}(k)$ in the interval $k\in[N_1,N_2]$, in which the average coupling in each KK level departs from zero. For even larger values, the average coupling decreases quickly, giving rise to very small contributions to the sum.

Physically, the intermediate region arises because, for fixed $\eta$, there exists a minimum KK mass $m_{q_n}\sim M_{\rm weak}/\eta$ beyond which the KK profile functions begin to penetrate the box modeling the Higgs-boson profile. When this happens, the cancellation of the Yukawa couplings of KK modes within one KK level is no longer operative. Only for yet much higher KK levels, when the profiles exhibit a large number of oscillations within the box, the couplings average out to zero and hence decrease with increasing $k$. While this behavior might seem strange at first sight, one should remember that by naive dimensional analysis the sum $\Sigma_q(N,\eta)$ is logarithmically divergent for large $N$. It is only due to subtle cancellations that it converges; however, when the profiles penetrate the box these cancellations no longer occur, and the generic  logarithmic growth arises.

\section{Relevance of the UV regulator}
\label{sec:cutoff}

In order to decide which of the two calculations presented in \cite{Casagrande:2010si} and \cite{Azatov:2010pf} is physically more meaningful, we recall the importance of using a consistent UV regularization scheme when evaluating the gluon-gluon fusion amplitude. This is true in the SM, and even more so in its 5D extensions. Here we will study two different regularization schemes: dimensional regularization and the use of a hard momentum cutoff. While a dimensional regulator is particular convenient in that it preserves gauge and Lorentz invariance, the second option is also a natural choice in the present case. This is because the RS model must be considered as an effective theory below the Planck scale, which requires a UV completion incorporating the effects of quantum gravity. A peculiar feature of warped extra-dimension models is that the effective UV cutoff depends on where the theory is probed along the extra dimension \cite{Randall:2001gb} (see also \cite{Pomarol:2000hp,Choi:2002wx,Goldberger:2002cz,Agashe:2002bx} for related works). The physical reason is that due to warping the fundamental length and energy scales change along the extra dimension. More specifically, the effective cutoff scale at the position $t$ in the extra dimension is of the order of the warped Planck scale
\begin{equation}
   \Lambda_{\rm UV}(t)\sim M_{\rm Pl}\,e^{-\sigma(\phi)} 
   = M_{\rm Pl}\,\frac{\epsilon}{t} \equiv \frac{\Lambda_{\rm TeV}}{t} \,. 
\end{equation}
The cutoff should be sufficiently high that at least a small number of KK modes have masses below $\Lambda_{\rm TeV}$, and hence $\Lambda_{\rm TeV}/M_{\rm KK}={\cal O}(10)$ or so. Otherwise incalculable ``threshold corrections'' of order $M_{\rm KK}^n/\Lambda_{\rm TeV}^n$ become important and take away the predictive power of the model. We stress that imposing a UV cutoff is crucial in order for the RS model to provide a viable solution to the hierarchy problem. Quantum corrections to the Higgs potential in the RS model exhibit an even stronger divergence than in the SM \cite{Casagrande:2008hr}. However, as long as the Higgs sector is localized near the IR brane (i.e., in the vicinity of $t=1$), the effective cutoff $\Lambda_{\rm UV}(t)$ is of the order of  several TeV, and hence the (big) hierarchy problem is solved.

The question of how to introduce such a cutoff in practical one-loop calculations is far from trivial. However, at large loop momentum (of order several times $M_{\rm KK}$) external momenta can be neglected, and hence there is a single 4D (euclidean) loop momentum $p_E^2\equiv-p^2$ on which the cutoff should be imposed. We propose to associate a $t$-dependent cutoff with every vertex of a Feynman diagram. This can be thought of as modeling the effect of a form factor, which accounts for  the impact of quantum gravity on energy scales above the effective Planck scale at that point. In general, the $t_i$ coordinates of the vertices are integrated over the entire bulk ($\epsilon\le t_i\le 1$), and hence the cutoff values vary between the TeV scale and the fundamental Planck scale. This is indeed an important effect, which makes gauge-coupling unification possible in warped extra-dimension models \cite{Randall:2001gb,Pomarol:2000hp,Choi:2002wx,Goldberger:2002cz,Agashe:2002bx}. The situation simplifies considerably for one-loop diagrams containing vertices with Higgs bosons, such as the one in Figure~\ref{fig:graph}. Denoting the coordinate of the two gluons by $t_1$ and $t_2$ and that of the Higgs boson by $t_3$, the fact that $t_3\approx 1$ ensures that the momentum cutoff on the 4D loop integral is
\begin{equation}\label{eq:cutoff}
   p_E \le \mbox{min}\,\big\{ \Lambda_{\rm UV}(t_1), \Lambda_{\rm UV}(t_2),
    \Lambda_{\rm UV}(t_3) \big\}
   = \Lambda_{\rm TeV} \,. 
\end{equation}
The same mechanism guarantees that the hierarchy problem is solved in RS models.

It is well known that evaluating the triangle diagram in Figure~\ref{fig:graph} in the 4D theory requires a consistent, gauge-invariant regulator such as dimensional regularization, the reason being that the integral is superficially UV divergent. If one were to evaluate it in $D=4$ dimensions, then a momentum-independent term proportional to $g^{\mu\nu}$ would appear, violating gauge invariance. In order to regularize the loop integral we thus evaluate it in $D=4-2\epsilon$ dimensions, with $\epsilon>0$, so as to regularize UV divergences. We then find that the sum in (\ref{Sigdef}) gets modified to
\begin{equation}
   \sum_{n=1}^{3+6N}\,\frac{v g_{nn}^q}{m_{q_n}} \left( \frac{\mu^2}{m_{q_n}^2} \right)^\epsilon ,
\end{equation}
where $\mu\sim\Lambda_{\rm TeV}$ is the regularization scale. For very large masses $m_{q_n}\gg\mu$, the dimensional regulator gives rise to a suppression, which renders the sum over KK modes convergent even for arbitrary ${\cal O}(1)$ Yukawa couplings. The limits $N\to\infty$ and $\eta\to 0$ can therefore be taken without encountering any ambiguities. The contribution from the dangerous intermediate region from super-massive KK modes with $m_{q_n}\sim M_{\rm weak}/\eta$ in Figure~\ref{fig:surprise}, which previously gave rise to an unsuppressed contribution of the form (\ref{loggrowth}), now receives an extra suppression factor $\eta^{2\epsilon}$, and vanishes when one takes the limit $\eta\to 0$ (at fixed $\epsilon$). The infinite sum then coincides with the result (\ref{ourS}) up to harmless ${\cal O}(\epsilon)$ corrections.

In the dimensional regularization scheme, gauge invariance is manifest in the 4D theory. Since the dimensional regulator also regularizes the infinite KK sum, we are guaranteed that the 5D theory remains gauge invariant, too. On the other hand, dimensional regularization is perhaps not the most intuitive way in which to introduce a UV cutoff. As an alternative, we will therefore rephrase the discussion in a regularization scheme based on using the hard momentum cutoff given in (\ref{eq:cutoff}). In order to ensure 4D gauge invariance in this case, we first extract two powers of the external gluon momenta by taking appropriate derivatives, after which the remaining loop integral is superficially convergent. Introducing the UV cutoff on this integral, and neglecting the Higgs-boson mass compared with $m_{q_n}$, we obtain
\begin{equation}\label{cutoffint}
   2v g_{nn}^q \int_0^{\Lambda_{\rm TeV}^2}\!dp_E^2\,p_E^2\,
    \frac{m_{q_n}}{\left(p_E^2+m_{q_n}^2\right)^3}
   = \frac{v g_{nn}^q}{m_{q_n}} 
    \left( \frac{\Lambda_{\rm TeV}^2}{\Lambda_{\rm TeV}^2+m_{q_n}^2} \right)^2 .
\end{equation}
For small masses $m_{q_n}\ll\Lambda_{\rm TeV}$ this reduces to the simple expression $v g_{nn}^q/m_{q_n}$ used in the sum (\ref{Sigdef}). For very large masses $m_{q_n}\gg\Lambda_{\rm TeV}$, on the other hand, the UV cutoff gives rise to a strong suppression proportional to $\Lambda_{\rm TeV}^4/m_{q_n}^4$, implying that such heavy KK modes decouple rapidly. It follows that, due to physical reasons, the sum over KK modes in (\ref{Sigdef}) is effectively truncated once the KK masses exceed the scale $\Lambda_{\rm TeV}$. 
In the RS model with a brane-localized Higgs sector, the scale $M_{\rm weak}/\eta$ at which the high-mass KK modes start to contribute a positive contribution to $\Sigma_q(N,\eta)$ with logarithmic growth is parametrically much larger than the effective cutoff scale $\Lambda_{\rm TeV}$. It is then appropriate to truncate the sum at a value $N_{\rm KK}^{\rm max}\sim\Lambda_{\rm TeV}/M_{\rm KK}$ corresponding to KK masses much smaller than $M_{\rm weak}/\eta$. It follows that
\begin{equation} \label{eq:threshold}
   \lim_{\eta\to 0}\,\Sigma_q(N_{\rm KK}^{\rm max},\eta)
   = \Sigma_q^{\rm (CGHNP)}
    + {\cal O}\bigg( \frac{N_{\rm KK}^{\rm max}\,v^2}{\Lambda_{\rm TeV}^2} \bigg) \,,    
\end{equation}
where the truncation error has the form of a threshold correction, which is always present in effective-theory calculations. 

We can summarize the above discussion by emphasizing the subtle fact that, in order to obtain the correct answer for the gluon-gluon fusion cross section in the RS model, it is essential to employ a consistent UV regularization scheme when evaluating the loop integral, despite of the fact that this integral is convergent. When this is done, the convergence of the infinite sum appearing in (\ref{doublelimit}) is improved in such a way that the order in which the two limits $N\to\infty$ and $\eta\to 0$ are taken becomes irrelevant. Specifically, in the two regularization schemes we have considered in our analysis, we find
\begin{equation}\label{twoschemes}
   \lim_{N\to\infty,~\eta\to 0}\,\sum_{n=1}^{3+6N} \frac{v g_{nn}^q}{m_{q_n}}
   \to \left\{ \begin{array}{l} \displaystyle \sum_{n=1}^\infty\,
    \frac{v g_{nn}^q}{m_{q_n}} \left( \frac{\mu^2}{m_{q_n}^2} \right)^\epsilon 
    \bigg|_{\eta=0} 
   \hspace{13.9mm} \!= \Sigma_q^{\rm (CGHNP)} + {\cal O}(\epsilon) \,, \\[6mm]
   \displaystyle \sum_{n=1}^\infty\,    
    \frac{v g_{nn}^q}{m_{q_n}} 
    \left( \frac{\Lambda_{\rm TeV}^2}{\Lambda_{\rm TeV}^2+m_{q_n}^2} \right)^2 
    \bigg|_{\eta=0} 
   \!= \Sigma_q^{\rm (CGHNP)}
    + {\cal O}\bigg( \frac{N_{\rm KK}^{\rm max}\,v^2}{\Lambda_{\rm TeV}^2} \bigg) \,.
   \end{array} \right.
\end{equation}
In both schemes the infinite sums are superficially convergent. The first line refers to the dimensional regularization scheme, while the second line corresponds to employing a hard momentum cutoff. It would be instructive to rederive these results by means of a properly UV-regularized 5D loop calculation, rather than by regularizing the sum over KK modes, as we have done above. In such a computation, in which there would be no reference to individual KK states, one should find that the correction to the $hgg$ amplitude is of the form (\ref{twoschemes}) once the regulator on the Higgs-boson profile is removed ($\eta\to 0$). For the simplified case of one generation, we have convinced ourselves that this is indeed the case. Details of this rather cumbersome analysis, as well as its extension to three generations, will be presented elsewhere.

The terms suppressed by a power of the UV cutoff, which appear on the right-hand side of (\ref{eq:threshold}), parameterize the difference between the asymptotic value $\Sigma_q^{\rm (CGHNP)}$ of the infinite sum and the sum over a finite number of KK modes. More generally, such threshold terms may also arise from the yet unknown effects of the UV completion of the RS model. From a low-energy perspective, the only requirement on such a completion that is relevant to us is that it must cure the hierarchy problem, by taming loop momenta exceeding the fundamental scale of quantum gravity. As long as this is the case, the gluon-gluon fusion amplitude will also be regularized in the way discussed above. In the context of a specific UV completion, the threshold effects could be modeled at low energies by means of a brane-localized effective $h\,G_{\mu\nu}^a\,G^{\mu\nu,a}$ operator, suppressed by $v/\Lambda_{\rm TeV}^2$. The additional factor $N_{\rm KK}^{\rm max}$ reflects the high multiplicity of degrees of freedom in the low-energy effective theory. For quite generic reasons, the coefficient of this operator must contain the loop factor $\alpha_s(\mu)/(4\pi)$ factored out in (\ref{Lhgg}), even in cases where the UV completion of the RS model is strongly coupled. The reason is that the on-shell external gluons couple proportional to their QCD charges, and that any new heavy state that couples to the Higgs boson must be color neutral, so it cannot have a tree-level coupling to gluons. Hence, a generic UV completion will indeed give rise to a threshold correction of the form shown in (\ref{eq:threshold}). The difference between the two sums (\ref{ourS}) and (\ref{theirS}), or the corresponding expressions (\ref{funny}) and (\ref{central_result}), can however {\em not\/} be attributed to such a brane-localized threshold term. Rather, it is related to the question whether a physical cutoff and a sensible UV completion are present at all.

\section{Phenomenology}
\label{sec:pheno}

We now study the implications of our results for Higgs-boson production in gluon-gluon fusion at the LHC. The master formula for the cross section has been given in (\ref{sigmaRS}), where the expressions for $\kappa_v$ and $\kappa_g$ can be found in (\ref{kappav}) and (\ref{kappas}), respectively, while to an excellent approximation we can set $\kappa_{g5}=0$. For the calculation of the Wilson coefficient $C_1^{\rm KK}$ we use our central result (\ref{central_result}) and assume that the parametrically-suppressed threshold effects appearing in (\ref{eq:threshold}) can be neglected. Note that the trace over a function $f(\bm{X}_q)$ of the matrix defined in (\ref{Xqdef}) is determined by the non-negative, real square roots $y_q^{(i)}$ of the eigenvalues of the hermitian matrices $\bm{Y}_q\,\bm{Y}_q^\dagger$, i.e.\
\begin{equation}\label{trace}
   \mbox{Tr}\,f(\bm{X}_q) 
   = \sum_i\,f\bigg(\frac{v\,y_q^{(i)}}{\sqrt2 M_{\rm KK}}\bigg) \,.
\end{equation}
Since in the RS model with anarchic 5D Yukawa couplings these matrices are structureless, the above result is proportional to the rank of the Yukawa matrices, which in the case of the minimal RS scenario is equal to the number of fermion generations. In other words, the KK towers of all six quarks give comparable contributions to the effective $hgg$ vertex, irrespective of the mass of the corresponding SM fermion. We emphasize that this feature is not present in many other extra-dimensional extensions of the SM. For instance, in models based on universal extra dimensions the 5D Yukawa couplings are hierarchical, like in the SM, and hence the Higgs-boson couplings to light fermions and their KK excitations are strongly suppressed \cite{Petriello:2002uu}. More interestingly, in several gauge-Higgs unification models, in which the Higgs appears as a pseudo-Goldstone boson, one finds that the contribution to the $hgg$ amplitude from the KK excitations of the SM quarks exactly cancels the dominant effect due to the corrections to the Yukawa couplings of the SM quarks, leaving only chirally-suppressed corrections, which are very small for all quarks other than the top quark \cite{Falkowski:2007hz}.\footnote{The authors of \cite{Azatov:2011qy} argue, however, that this cancellation is model dependent and depends on the embedding of the fermions in the composite multiplets.}  
Hence, in these new-physics scenarios only the top quark and its heavy partners contribute to the effective $hgg$ couplings, while Higgs-boson production is independent of the masses and couplings of the KK excitations of light SM quarks. 

By randomly generating a large set of 5D Yukawa matrices, which are required to satisfy $|(\bm{Y}_q)_{ij}|\le y_{\rm max}$ and to correctly reproduce the Wolfenstein parameters $\rho$ and $\eta$ of the unitarity triangle, we have found that to a good approximation the average result for the trace appearing in (\ref{trace}) can be parametrized as\footnote{In \cite{Goertz:2011hj}, we preferred to use the associated Yukawa matrices $\tilde{\bm{Y}}_q$ defined in (\ref{Xqdef}) instead of the original 5D Yukawa matrices $\bm{Y}_q$. In practice, it turns out to be immaterial whether one considers $\bm{Y}_q$ or $\tilde{\bm{Y}}_q$ as random complex matrices.}  
\begin{equation}\label{param1} 
   \mbox{Tr}\,f(\bm{X}_q)\approx 3 \left( 1 - 1.3\,x_{\rm max}^2 \right) f(x_{\rm max}) \,; 
    \qquad
   \mbox{with} \quad
   f(x) = \frac{x\tanh x}{\cosh 2x} \,,
\end{equation}
and $x_{\rm max}\equiv v\,y_{\rm max}/(\sqrt{2} M_{\rm KK})$. Likewise, the parameter $\kappa_t$ entering (\ref{kappas}) can be approximated as 
\begin{equation}\label{param2}
   \kappa_t\approx 1 - 1.1\,x_{\rm max}^2 - \varepsilon_u \,, 
\end{equation}
where to an excellent approximation the contribution proportional to $\varepsilon_u$ cancels in the sum $C_1^{\rm KK}(m_h)+{\rm Re}\left(\kappa_t\right) A(\tau_t)$ appearing in the numerator of the expression for $\kappa_g$ in (\ref{kappas}). Furthermore, the very small effects due to $\kappa_b$ and $\varepsilon_d$ can be neglected for all practical purposes.

\begin{figure}
\begin{center}
\includegraphics[height=0.45\textwidth]{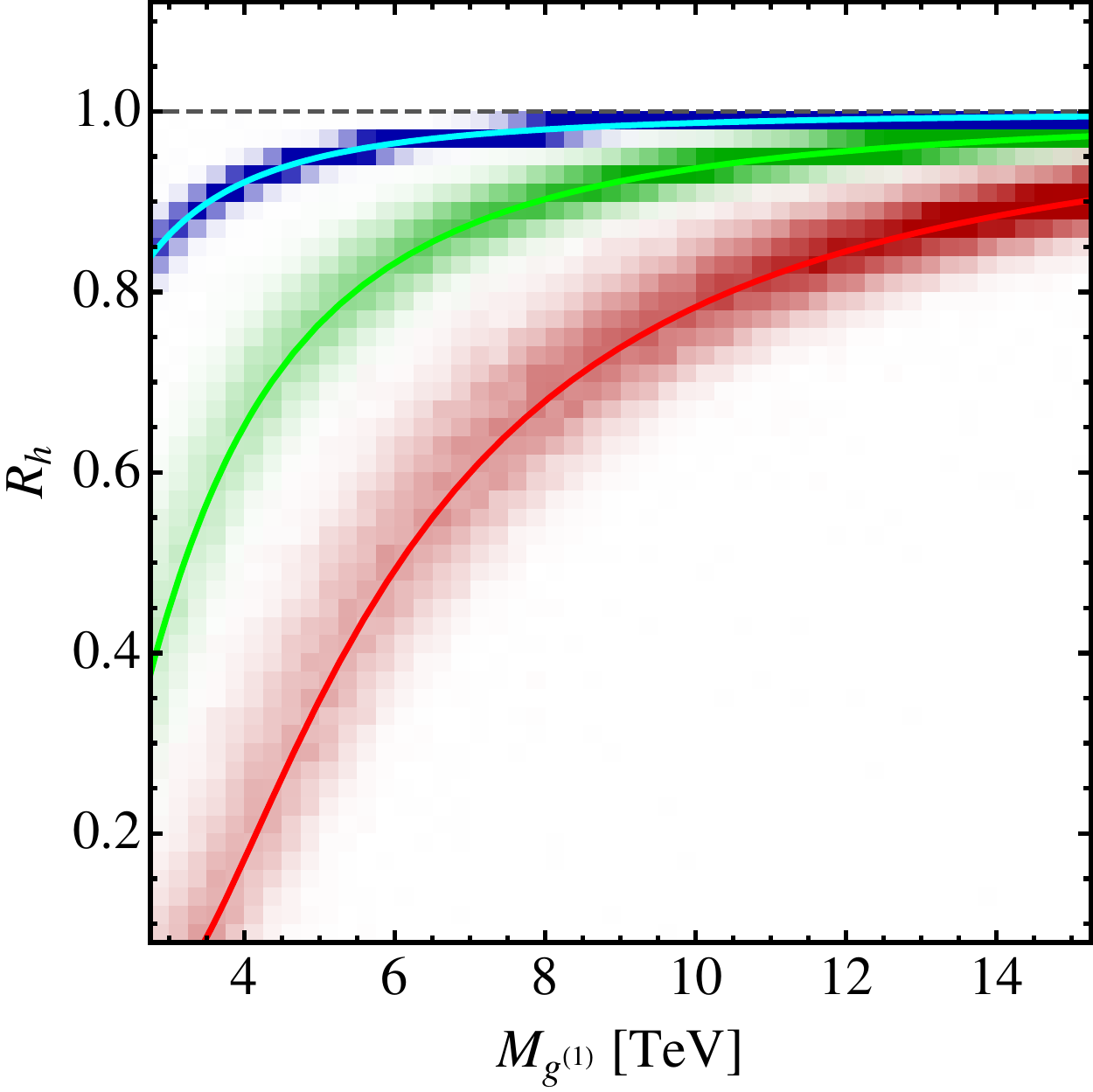}
\quad
\includegraphics[height=0.459\textwidth]{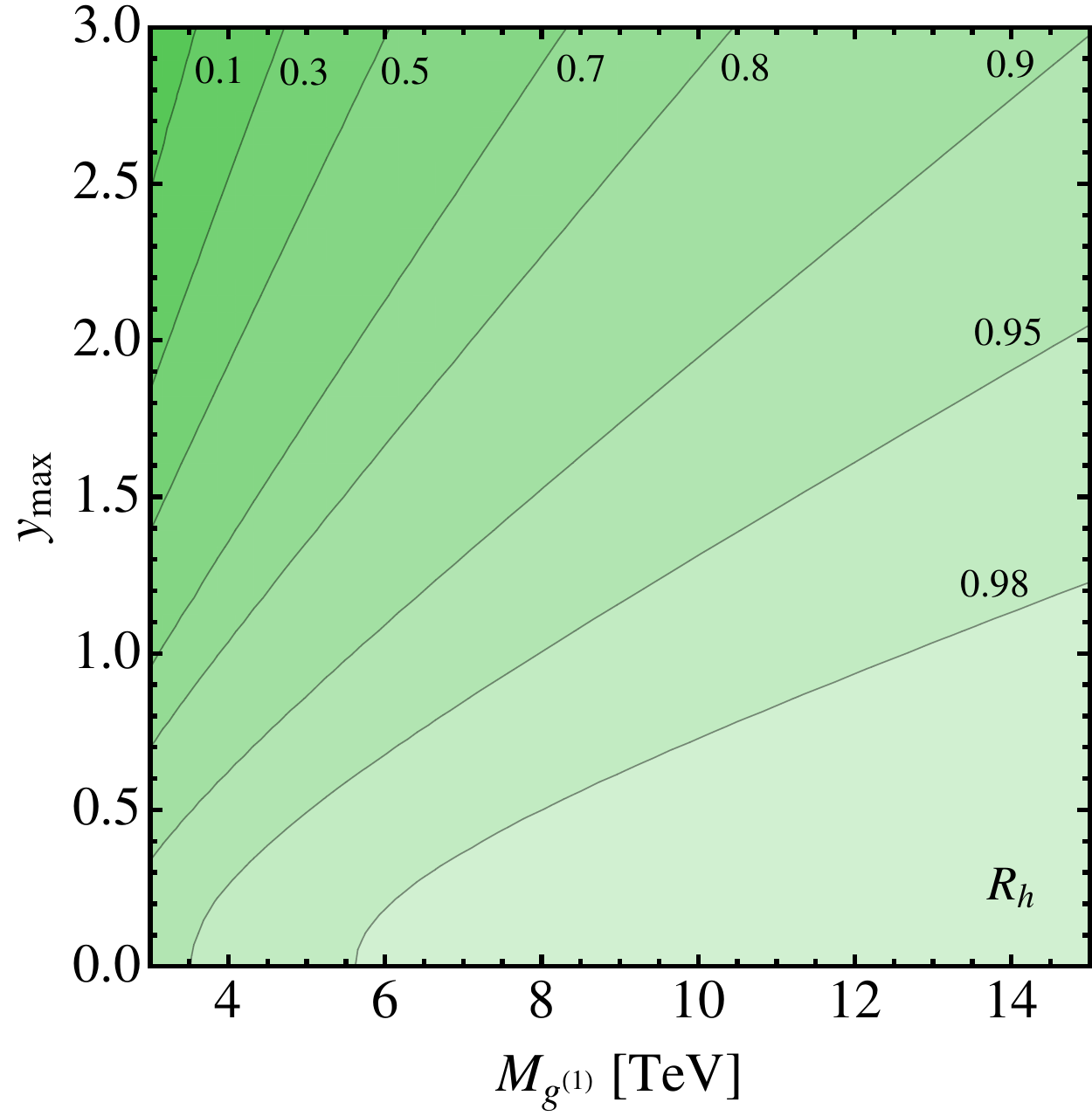} 
\parbox{15.5cm}
{\caption{\label{fig:Rh} 
{\em Left:} 
Predictions for the ratio $R_h$ in the minimal RS model with bulk matter fields and an IR-localized Higgs sector. The red, green, and blue density bands correspond to $y_{\rm max}=3$, 1.5, and 0.5, respectively. The overlaid solid lines are obtained using the approximate parameterizations given in (\ref{param1}) and (\ref{param2}) for the same values of $y_{\rm max}$. 
{\em Right:} 
Contour plot for the ratio $R_h$ obtained using the latter parameterizations.}}
\end{center}
\end{figure}

In the left plot in Figure~\ref{fig:Rh}, we show our results for the Higgs-boson production cross section in gluon-gluon fusion  relative to the SM cross section,
\begin{equation}
   R_h = \frac{\sigma(gg\to h)_{\rm RS}}{\sigma(gg\to h)_{\rm SM}} \,,
\end{equation}
as a function of the mass of the lightest KK gluon state, $M_{g^{(1)}}\approx 2.45\,M_{\rm KK}$. We use the lightest KK gluon mass as a reference, because its value is a model-independent prediction of the RS scenario. The masses of the lightest KK fermions have very similar values but depend to some extent on the bulk mass parameters. Here and below we employ a Higgs-boson mass of 125\,GeV. The solid red, green, and light blue lines show the approximate results obtained from (\ref{param1}) and (\ref{param2}) employing the values $y_{\rm max}=3$, 1.5, and 0.5, respectively. The underlaid density plots indicate the distribution of the predictions for a large number of anarchic Yukawa matrices $\bm{Y}_q$, subject only to the constraint that their elements are smaller in magnitude than a given value $y_{\rm max}$. The fact that the regions of highest density in the scatter plots nicely reproduce the results obtained directly from (\ref{param1}) and (\ref{param2}) shows that the requirement to reproduce the correct values of $\rho$ and $\eta$ does not play an important role numerically. The contour plot displayed on the right in the figure gives a two-dimensional representation for the cross section as a function of $M_{g^{(1)}}$ and $y_{\rm max}$, obtained by employing again (\ref{param1}) and (\ref{param2}).

We observe from Figure~\ref{fig:Rh} that $R_{h}$ is strictly below 1 and decreases (increases) with increasing $y_{\rm max}$ (KK scale). In other words, the minimal RS model with a brane-localized Higgs sector predicts a depletion of $\sigma(gg\to h)$ relative to the SM. In the region where $v\,y_{\rm max}/M_{\rm KK}$ is a suitable expansion parameter, we obtain the approximate result 
\begin{equation}
   R_h\approx 1 - \frac{v^2}{2M_{\rm KK}^2}\,\big( 14.2\,y_{\rm max}^2 + 3.5 \big) \,, 
\end{equation}
where the constant term in parenthesis is due to the effect of $\kappa_v$. For not too small Yukawa couplings this observable is dominated by the effects of KK quark loops. Given the strong dependence of the ratio $R_h$ on $y_{\rm max}$, we find that for $M_g^{(1)}\approx 3$\,TeV and Yukawa couplings close to the perturbativity bound $y_{\rm max}\approx 3$ \cite{Csaki:2008zd}, the new-physics contributions to the Higgs-boson production cross section in gluon-gluon fusion can become so large that they completely cancel the SM contribution. In fact, the sensitivity of $R_h$ to the overall size of the 5D Yukawa couplings is even more pronounced than the one arising in the case of dipole-operator transitions such as $B\to X_s\gamma$ \cite{Agashe:2008uz}. While the latter contributions also scale with $y_{\rm max}^2$, unlike $R_h$ they are (at the one-loop level) insensitive to the multiplicity of states in the fermionic sector of the RS model under consideration. This feature underscores our assertion (made in the introduction) that precision measurements of the Higgs-boson properties furnish a superb tool for illuminating the quantum structure of electroweak interactions in RS scenarios. 

In the minimal RS model considered here, constraints from electroweak precision observables \cite{Carena:2003fx} and flavor physics \cite{Csaki:2008zd} require that the lightest KK excitations of SM particles must have masses in the 10\,TeV range, which puts them outside of the reach for production at the LHC. Figure~\ref{fig:Rh} shows that even in this case there can be significant virtual effects of KK particles on the Higgs-boson production cross section, provided that the 5D Yukawa couplings are not too small. The bounds from electroweak precision measurements, in particular, can be relaxed in several ways. For instance, ``little RS models'', in which the size $L$ of the extra dimension is reduced \cite{Davoudiasl:2008hx}, would only have a minor impact on our analysis. Other extensions, such as models with a custodial $SU(2)_R$ gauge symmetry in the bulk \cite{Agashe:2003zs,Agashe:2006at}, might however give rise to a rather different Higgs-boson phenomenology \cite{Casagrande:2010si,Goertz:2011hj}.

\section{Conclusions}
\label{sec:concl}

The announcements of the first direct hints for a Higgs-boson signal by the LHC and Tevatron experiments open up a new chapter in particle physics. Although the significance of the various measurements is not yet sufficient to preclude the possibility of statistical fluctuations accounting for the observed effects, hopes are high that, with ATLAS and CMS accumulating more data, a Higgs boson will be discovered (or excluded) by the end of this year. This discovery would not only mark the birth of the hierarchy problem, but it will also reshape some of the fundamental questions of our field. In particular, the focus of large parts of the LHC physics program will shift towards determining the Higgs-boson properties as accurately as possible, with the ultimate goal of probing possible deviations from the SM expectations. In close analogy to flavor physics, precision Higgs-boson physics represents a powerful way to investigate the dynamics of electroweak symmetry breaking at the quantum level. In fact, the couplings of the Higgs boson to a pair of gluons and photons vanish at tree level in the SM, but are induced by the exchange of virtual top quarks and $W$ bosons at one-loop order. The effective $hgg$ and $h\gamma\gamma$ couplings thus offer a distinctive window to physics beyond the SM, where new heavy particles can propagate in the loops, thereby potentially affecting both the production cross section and the decay rates of the Higgs boson. 

The main goal of this article was to perform an analytic calculation of the impact of KK fermions on the production of a SM-like Higgs boson in the minimal RS scenario featuring an $SU(2)_L\times U(1)_Y$ bulk gauge symmetry and a brane-localized Higgs sector. In particular, we have revisited the gluon-gluon fusion process, for which two independent calculations previously found contradictory results. A significant suppression of the cross section was reported in \cite{Casagrande:2010si}, while in \cite{Azatov:2010pf} an effect of similar magnitude but opposite sign was obtained. As we have shown, the discrepancy is not due to a simple computational mistake. On a technical level, it can be traced to the fact that, in order to make the overlap integrals of the wave functions of the brane-localized Higgs-boson with the bulk fermions mathematically well defined, a regularization of the Higgs profile is unavoidable in an intermediate step of the calculation. It is achieved by smearing out the profile over a finite width $\eta$, thereby moving the Higgs boson slightly into the bulk. Given that the Yukawa couplings depend on $\eta$, and that the calculations in \cite{Casagrande:2010si,Azatov:2010pf} both involve the summation over an infinite number $N$ of KK levels, the question arises whether the result for the $gg\to h$ amplitude might depend on the order in which the limits $\eta\to 0$ and $N\to\infty$ are taken. We have demonstrated that the order in which these limits are performed indeed explains the aforementioned discrepancy. While the results of \cite{Casagrande:2010si} are reproduced if one first takes the limit of the Higgs regulator to zero and then sums over the infinite tower of KK states, the findings of \cite{Azatov:2010pf} correspond to the reversed order of taking the two limits. 

We have presented for the first time closed analytical expressions for both results in terms of the fundamental parameters of the RS model, valid to all orders in the ratio $v/M_{\rm KK}$ of the Higgs vacuum expectation value and the KK mass scale. We have then pointed out that the non-commutativity of the limits $\eta\to 0$ and $N\to\infty$ is due to a hidden UV sensitivity of the infinite KK sum, and illustrated this phenomenon by means of a simple toy model, which can be solved exactly. This computation highlights that for fixed $\eta$ there exists a region, starting at around $N\approx 0.1/\eta$, beyond which the KK-fermion wave functions resolve the width of the Higgs profile. As a result, the subtle cancellation of the Yukawa couplings of KK modes within each KK level, which is instrumental for the convergence of the KK sum, is spoiled. Thus logarithmic growth of the $gg\to h$ amplitude kicks in, which continues until the KK fermion profiles exhibit a large number of oscillations within the region where the Higgs profile is localized. Excitations with KK numbers higher than $N\approx 10/\eta$ quickly decouple, rendering the infinite KK sum ultimately convergent. 

The question which of the two calculations of the gluon-gluon fusion amplitude gives the correct physics result is resolved by introducing a proper UV regularization scheme. Specifically, we have studied the problem using dimensional regularization and using a hard momentum cutoff, which may be identified with the inherent UV cutoff of RS models set by the warped Planck scale. In both cases, the regulator improves the convergence of the KK sum and removes the unphysical, previously unsuppressed contributions from super-heavy KK modes. The regularized sum over KK states yields the result of \cite{Casagrande:2010si}, up to possible threshold corrections suppressed by inverse powers of the effective UV cutoff. This shows that Higgs production is an UV-insensitive process in the RS model and can be calculated unambiguously once a proper regularization is employed. It would be worthwhile to reproduce our main result (\ref{central_result}) for the induced $hgg$ vertex by performing an actual 5D loop calculation rather than an infinite KK sum. This is left for future work.

We have finally studied the numerical impact of our results for Higgs-boson production in gluon-gluon fusion, thereby extending our previous, more phenomenologically oriented analyses of Higgs physics in RS models \cite{Casagrande:2010si,Goertz:2011hj}. We have reemphasized the important point that, regardless of the mass of the corresponding SM fermion, in warped extra-dimensional models the loop-induced couplings of the Higgs field to gauge bosons receive similar contributions from the KK towers of each fermion state, and that these contributions scale with the square of the 5D Yukawa couplings. Assuming a similar KK mass scale, RS scenarios hence predict much stronger effects than many other extra-dimensional extensions of the SM, such as universal extra dimensions and models in which the Higgs emerges as a pseudo-Goldstone boson. While we believe this fact to be true in general, it is not unlikely that the structure of the corrections depends on the precise realization of the scalar sector. Exploring the associated model dependence in more detail is left for future work. In order to probe the large effects predicted by our calculations, it would be particularly useful to measure the cross sections for Higgs-boson production in both the gluon-gluon and vector-boson fusion channels. A combination of theses two measurements would allow for a clean extraction of the $gg\to h$ amplitude, and hence represents a unique way to study the Yukawa sector of RS theories.

\vspace{3mm}
{\em Acknowledgements:\/}
We are grateful to Kaustubh Agashe, Aleksandr Azatov, Adrian Carmona, Lisa Randall, Manuel Perez-Victoria, Edurado Ponton, Jose Santiago, and Fabio Zwirner for useful discussions. The research of S.C.\ is supported by the DFG Cluster of Excellence {\em Origin and Structure of the Universe}. The research of F.G.\ is supported by the Swiss National Foundation under contract SNF 200020-126632. The research of M.N.\ is supported by the Advanced Grant EFT4LHC of the European Research Council (ERC), grant 05H09UME of the German Federal Ministry for Education and Research (BMBF), and the Rhineland-Palatinate Research Center {\em Elementary Forces and Mathematical Foundations}. Fermilab is operated by Fermi Research Alliance, LLC under Contract No. DE-AC02-07CH11359 with the U.S.\ Department of Energy.

\begin{appendix}

\section{Special case of vanishing bulk masses}
\label{sec:toy}
\renewcommand{\theequation}{A\arabic{equation}}
\setcounter{equation}{0}

In the case where all bulk mass parameters $c_A$ are set to zero, and where in addition we set $\epsilon=0$ (which in this case is unproblematic), the EOMs (\ref{eoms}) can be solved exactly if we work with a sufficiently simple regularized Higgs profile. This provides a nice test case for our general results and conclusions. In order to obtain analytic expressions for the wave functions of the individual KK modes, we consider the Higgs profile (\ref{Hbox}) and furthermore assume that there is a single fermion generation with real and positive Yukawa coupling $Y_q$. The generalized mass matrix in (\ref{Mqdef}) is then given by 
\begin{equation}
   {\cal M}_q(t) = \begin{cases}\displaystyle 
    \; \frac{v Y_q}{\sqrt2 M_{\rm KK}}\,\frac{1}{\eta}\, 
     \bigg( \begin{array}{cc} 0~ & 1 \\ 1~ & 0 \end{array} \bigg)
     \equiv z\,\bigg( \begin{array}{cc} 0~ & 1 \\ 1~ & 0 \end{array} \bigg) \,; 
     & \text{for \;}  1-\eta < t < 1 \,, \\[4mm] 
    \hspace{3.0cm} 0 \,; & \text{for \;} \epsilon \leq t < 1-\eta \,. 
    \end{cases}
\end{equation}
The wave functions ${\cal Q}_A^{(n)}(t)$ are now 2-component objects. 

In the solution of the EOMs we must distinguish two cases. For $x_n<z$, we find
\begin{equation}
   \left( \begin{array}{c} {\cal Q}_L^{(n)}(t) \\ {\cal Q}_R^{(n)}(t) \end{array} \right)
   = N_n \left( \begin{array}{c} \cos(x_n t) \\ \mp\sin(x_n t) \\ 
    \sin(x_n t) \\ \pm\cos(x_n t) \end{array} \right) \,; 
   \quad t\le 1-\eta \,,
\end{equation}
and
\begin{equation}
   \left( \begin{array}{c}{\cal Q}_L^{(n)}(t) \\ {\cal Q}_R^{(n)}(t) \end{array} \right)
   = N_n \left( \begin{array}{c} r_1 \cosh\big[\sqrt{z^2-x_n^2}\,(1-t)\big] \\ 
    \mp r_2\,\sinh\big[\sqrt{z^2-x_n^2}\,(1-t)\big] \\ 
    r_2\,\sinh\big[\sqrt{z^2-x_n^2}\,(1-t)\big] \\
    \pm r_1 \cosh\big[\sqrt{z^2-x_n^2}\,(1-t)\big] \end{array} \right) \,; 
   \quad t\ge 1-\eta \,,
\end{equation}
where the coefficients $r_i$ are given by
\begin{equation} 
   r_1 = \frac{\cos\big[x_n(1-\eta)\big]}{\cosh\big[\sqrt{z^2-x_n^2}\,\eta\big]} \,, \qquad
   r_2 = \frac{\sin\big[x_n(1-\eta)\big]}{\sinh\big[\sqrt{z^2-x_n^2}\,\eta\big]} \,.
\end{equation}
For $x_n>z$, we find instead
\begin{equation}
   \left( \begin{array}{c} {\cal Q}_L^{(n)}(t) \\ {\cal Q}_R^{(n)}(t) \end{array} \right)
   = N_n \left( \begin{array}{c} r_1 \cos\big[\sqrt{x_n^2-z^2}\,(1-t)\big] \\ 
    \mp r_2\,\sin\big[\sqrt{x_n^2-z^2}\,(1-t)\big] \\ 
    r_2\,\sin\big[\sqrt{x_n^2-z^2}\,(1-t)\big] \\
    \pm r_1 \cos\big[\sqrt{x_n^2-z^2}\,(1-t)\big] \end{array} \right) \,; 
   \quad t\ge 1-\eta \,,
\end{equation}
where
\begin{equation}
   r_1 = \frac{\cos\big[x_n(1-\eta)\big]}{\cos\big[\sqrt{x_n^2-z^2}\,\eta\big]} \,, \qquad
   r_2 = \frac{\sin\big[x_n(1-\eta)\big]}{\sin\big[\sqrt{x_n^2-z^2}\,\eta\big]} \,.
\end{equation}

The eigenvalues $x_n$ are determined by matching the profile functions at $t=1-\eta$. This leads to the conditions shown in (\ref{eigenvals}). Evaluating the normalization constraint, we obtain the expression for $N_n$ given in the second line of (\ref{toysum2}). Using these results, it is not difficult to calculate the relevant overlap integrals with the Higgs profile. In this way we arrive at the result for the sum $\Sigma_q(N,\eta)$ shown in the first line of (\ref{toysum2}).

\end{appendix}


\newpage
\begin{thebibliography}{99}

\bibitem{Randall:1999ee}  
  L.~Randall and R.~Sundrum,  
  Phys.\ Rev.\ Lett.\  {\bf 83}, 3370 (1999)  
  [hep-ph/9905221].  

\bibitem{Grossman:1999ra}
  Y.~Grossman and M.~Neubert,
  Phys.\ Lett.\ B {\bf 474}, 361 (2000)
  [hep-ph/9912408].

\bibitem{Gherghetta:2000qt}
  T.~Gherghetta and A.~Pomarol,
  Nucl.\ Phys.\ B {\bf 586}, 141 (2000)
  [hep-ph/0003129].

\bibitem{Huber:2000ie}
  S.~J.~Huber and Q.~Shafi,
  Phys.\ Lett.\ B {\bf 498}, 256 (2001)
  [hep-ph/0010195].

\bibitem{Agashe:2004ay} 
  K.~Agashe, G.~Perez and A.~Soni,
  Phys.\ Rev.\ Lett.\  {\bf 93}, 201804 (2004)
  [hep-ph/0406101].

\bibitem{Agashe:2004cp} 
  K.~Agashe, G.~Perez and A.~Soni,
  Phys.\ Rev.\ D {\bf 71}, 016002 (2005)
  [hep-ph/0408134].

\bibitem{Davoudiasl:2002ua} 
  H.~Davoudiasl, J.~L.~Hewett and T.~G.~Rizzo,
  Phys.\ Rev.\ D {\bf 68}, 045002 (2003)
  [hep-ph/0212279].

\bibitem{Carena:2002dz} 
  M.~S.~Carena, E.~Ponton, T.~M.~P.~Tait and C.~E.~M.~Wagner,
  Phys.\ Rev.\ D {\bf 67}, 096006 (2003)
  [hep-ph/0212307].

\bibitem{Agashe:2003zs}
  K.~Agashe, A.~Delgado, M.~J.~May and R.~Sundrum,
  JHEP {\bf 0308}, 050 (2003)
  [hep-ph/0308036].

\bibitem{Agashe:2006at}
  K.~Agashe, R.~Contino, L.~Da Rold and A.~Pomarol,
  Phys.\ Lett.\ B {\bf 641}, 62 (2006)
  [hep-ph/0605341].
  
\bibitem{Bauer:2011ah} 
  M.~Bauer, R.~Malm and M.~Neubert,
  Phys.\ Rev.\ Lett.\  {\bf 108}, 081603 (2012)
  [arXiv:1110.0471 [hep-ph]].

\bibitem{Contino:2003ve}
  R.~Contino, Y.~Nomura and A.~Pomarol,
  Nucl.\ Phys.\ B {\bf 671}, 148 (2003)
  [hep-ph/0306259].
   
\bibitem{Agashe:2004rs}
  K.~Agashe, R.~Contino and A.~Pomarol,
  Nucl.\ Phys.\ B {\bf 719}, 165 (2005)
  [hep-ph/0412089].

\bibitem{Casagrande:2010si}
  S.~Casagrande, F.~Goertz, U.~Haisch, M.~Neubert and T.~Pfoh,
  JHEP {\bf 1009}, 014 (2010)
  [arXiv:1005.4315 [hep-ph]].
 
\bibitem{Azatov:2010pf}
  A.~Azatov, M.~Toharia and L.~Zhu,
  Phys.\ Rev.\  D {\bf 82}, 056004 (2010)
  [arXiv:1006.5939 [hep-ph]].

\bibitem{Davoudiasl:1999tf} 
  H.~Davoudiasl, J.~L.~Hewett and T.~G.~Rizzo,
  Phys.\ Lett.\ B {\bf 473}, 43 (2000)
  [hep-ph/9911262].

\bibitem{Huber:2003tu} 
  S.~J.~Huber,
  Nucl.\ Phys.\ B {\bf 666}, 269 (2003)
  [hep-ph/0303183].
  
\bibitem{Csaki:2008zd}
  C.~Csaki, A.~Falkowski and A.~Weiler,
  JHEP {\bf 0809}, 008 (2008)
  [arXiv:0804.1954 [hep-ph]].
  
\bibitem{Casagrande:2008hr}
  S.~Casagrande, F.~Goertz, U.~Haisch, M.~Neubert and T.~Pfoh,
  JHEP {\bf 0810}, 094 (2008)
  [arXiv:0807.4937 [hep-ph]].
 
\bibitem{Blanke:2008zb} 
  M.~Blanke, A.~J.~Buras, B.~Duling, S.~Gori and A.~Weiler,
  JHEP {\bf 0903}, 001 (2009)
  [arXiv:0809.1073 [hep-ph]].

\bibitem{Azatov:2009na} 
  A.~Azatov, M.~Toharia and L.~Zhu,
  Phys.\ Rev.\ D\ {\bf 80}, 035016  (2009)
  [arXiv:0906.1990 [hep-ph]].

\bibitem{Bouchart:2009vq} 
  C.~Bouchart and G.~Moreau,
  Phys.\ Rev.\ D {\bf 80}, 095022 (2009)
  [arXiv:0909.4812 [hep-ph]].

\bibitem{Beneke:2002jn} 
  M.~Beneke and M.~Neubert,
  Nucl.\ Phys.\ B\ {\bf 651}, 225  (2003)
  [hep-ph/0210085].

\bibitem{Djouadi:2005gj} 
  A.~Djouadi,
  Phys.\ Rept.\  {\bf 459}, 1 (2008)
  [hep-ph/0503173]. 

\bibitem{Goertz:2011hj}
  F.~Goertz, U.~Haisch and M.~Neubert,
  Phys.\ Lett.~B {\bf 713}, 23 (2012) [arXiv:1112.5099 [hep-ph]].

\bibitem{Inami:1982xt} 
  T.~Inami, T.~Kubota and Y.~Okada,
  Z.\ Phys.\ C {\bf 18}, 69 (1983).

\bibitem{Ahrens:2008nc} 
  V.~Ahrens, T.~Becher, M.~Neubert and L.~L.~Yang,
  Eur.\ Phys.\ J.\ C\ {\bf 62}, 333  (2009)
  [arXiv:0809.4283 [hep-ph]].

\bibitem{Puchwein:2003jq} 
  M.~Puchwein and Z.~Kunszt,
  Annals Phys.\  {\bf 311}, 288 (2004)
  [hep-th/0309069].
  
\bibitem{Carena:2004zn} 
  M.~S.~Carena, A.~Delgado, E.~Ponton, T.~M.~P.~Tait and C.~E.~M.~Wagner,
  Phys.\ Rev.\ D {\bf 71}, 015010 (2005)
  [hep-ph/0410344].
  
\bibitem{Falkowski:2007hz}
  A.~Falkowski,
  Phys.\ Rev.\  D {\bf 77}, 055018 (2008)
  [arXiv:0711.0828 [hep-ph]].

\bibitem{Randall:2001gb} 
  L.~Randall and M.~D.~Schwartz,
  JHEP {\bf 0111}, 003 (2001)
  [hep-th/0108114].

\bibitem{Pomarol:2000hp} 
  A.~Pomarol,
  Phys.\ Rev.\ Lett.\  {\bf 85}, 4004 (2000)
  [hep-ph/0005293].
  
\bibitem{Choi:2002wx} 
  K.~-w.~Choi, H.~D.~Kim and I.~-W.~Kim,
  JHEP {\bf 0211}, 033 (2002)
  [hep-ph/0202257].

\bibitem{Goldberger:2002cz} 
  W.~D.~Goldberger and I.~Z.~Rothstein,
  Phys.\ Rev.\ Lett.\  {\bf 89}, 131601 (2002)
  [hep-th/0204160].
  
\bibitem{Agashe:2002bx} 
  K.~Agashe, A.~Delgado and R.~Sundrum,
  Nucl.\ Phys.\ B {\bf 643}, 172 (2002)
  [hep-ph/0206099].

\bibitem{Petriello:2002uu} 
  F.~J.~Petriello,
  JHEP {\bf 0205}, 003 (2002)
  [hep-ph/0204067].

\bibitem{Azatov:2011qy} 
  A.~Azatov and J.~Galloway,
  Phys.\ Rev.\ D {\bf 85}, 055013 (2012)
  [arXiv:1110.5646 [hep-ph]].

\bibitem{Agashe:2008uz}
  K.~Agashe, A.~Azatov and L.~Zhu,
  Phys.\ Rev.\  D {\bf 79}, 056006 (2009)
  [arXiv:0810.1016 [hep-ph]].

\bibitem{Carena:2003fx} 
  M.~S.~Carena, A.~Delgado, E.~Ponton, T.~M.~P.~Tait and C.~E.~M.~Wagner,
  Phys.\ Rev.\ D {\bf 68}, 035010 (2003)
  [hep-ph/0305188].

\bibitem{Davoudiasl:2008hx} 
  H.~Davoudiasl, G.~Perez and A.~Soni,
  Phys.\ Lett.\ B {\bf 665}, 67 (2008)
  [arXiv:0802.0203 [hep-ph]].
  
\end{thebibliography}
\end{document}